\documentclass[A4]{aa}
\usepackage{txfonts}
\usepackage{graphicx}
\def\M21{\hbox{Mrk\,421} } \def\xmm{\hbox{$XMM-Newton$}} \def\etal{et
  al. } \def\ie{i.e., } 
 \def\ltsima{$\; \buildrel < \over
  \sim \;$} \def\simlt{\lower.5ex\hbox{\ltsima}} % < over ~
\def\gtsima{$\; \buildrel > \over \sim \;$}
\def\simgt{\lower.5ex\hbox{\gtsima}} % > over ~
\newcommand{\eg}{e.g.,\ }
\newcommand{\eqref}[1]{(\ref{#1})}
\newcommand{\dsfrac}[2]{\displaystyle{\frac{#1}{#2}}} \def\difd{{\rm
    d}}   

\def\ab{\alpha_{\rm B}}
\def\ae{\alpha_\mathrm{e}}
\newcommand{\SF}[2]{{\bf S}#1-{\bf F}#2}
\def\pmag{p_{\rm mag}} 
\def\epsth{\epsilon_{\rm th}} 
\def\epsmag{\epsilon_{\rm mag}} 
\def\gad{\gamma_{\rm ad}}
\def\BP{B^{\prime}}
\newcommand{\BG}[1]{{\bf B}#1} 
\newcommand{\TG}[1]{{\bf T}#1}

\begin{document}

\title{Internal shocks in relativistic outflows:\\ collisions of
       magnetized shells}
\author{P.\,Mimica\inst{1}, M.A.\,Aloy\inst{1,2} and E.\,M\"uller\inst{2}}
\offprints{PM, e-mail: mimica@uv.es}
\institute{Departamento de Astronom\'{\i}a y Astrof\'{\i}sica, 
           Universidad de Valencia, 46100 Burjassot, Spain
\and       Max-Planck-Institut f\"ur Astrophysik,
           Postfach 1312, D-85741 Garching, Germany}
\date{Received ?; accepted ?}
\abstract
    {}
% aims (obligatory)
    { We study the collision of magnetized irregularities (shells) in
      relativistic outflows in order to explain the origin of the
      generic phenomenology observed in the non-thermal emission of
      both blazars and gamma-ray bursts. We focus on the influence of
      the magnetic field on the collision dynamics, and we investigate
      how the properties of the observed radiation depend on the
      strength of the initial magnetic field and on the initial
      internal energy density of the flow .}
% method (obligatory )
    { The collisions of magnetized shells and the radiation resulting
      from these collisions are calculated using the 1D relativistic
      magnetohydrodynamics code \emph{MRGENESIS}. The interaction of
      the shells with the external medium prior to their collision is
      also determined using an exact solver for the corresponding 1D
      relativistic magnetohydrodynamic Riemann problem. In both cases
      we assume that the magnetic field is oriented perpendicular to
      the flow direction. }
% results (obligatory)
    { Our simulations show that two magnetization parameters - the
      ratio of magnetic energy density and thermal energy density,
      $\alpha_B$, and the ratio of magnetic energy density and
      mass-energy density, $\sigma$ - play an important role in the
      pre-collision phase, while the dynamics of the collision and the
      properties of the light curves depend mostly on the
      magnetization parameter $\sigma$. Comparing synthetic light
      curves computed from hydrodynamic and magnetohydrodynamic models
      we find that the assumption commonly made in the former models
      that the magnetization parameter $\alpha_B$ is constant and
      uniform, holds rather well, if $\alpha_B < 0.01$. The
      interaction of the shells with the external medium changes the
      flow properties at their edges prior to the collision. For
      sufficiently dense shells moving at large Lorentz factors
      ($\simgt 25$) these properties depend only on the magnetization
      parameter $\sigma$. Internal shocks in GRBs may reach maximum
      efficiencies of conversion of kinetic into thermal energy
      between $6\%$ and $10\%$, while in case of blazars, the maximum
      efficiencies are $\sim 2\%$.} 
{}

\titlerunning{Collisions of magnetized shells} 
\authorrunning{Mimica et al.}  \maketitle

\section{Introduction} 
  \label{intro}
  
  Relativistic outflows have been observed extensively in blazars, a
  class of active galactic nuclei (AGN) known to show the most rapid
  variability of all AGNs. Their remarkable characteristic, flares
  in the X-ray frequency range, usually have a duration  of the order
  of one day (\cite{MA99}\,1999, \cite{TA00}\,2000). With the improved
  sensitivity of \xmm, variability on time scales of kilo-seconds has
  been studied in the source Mrk 421 (\cite{BR01}\,2001, 2003;
  \cite{R04}\,2004), and recently the spectral evolution of this
  object down to time scales of $\approx 100$\,s could be followed
  (\cite{BR05}\,2005).
  
  Often, the internal shock scenario (\cite{RM94}\,1994) is invoked to
  explain the variability of blazars (see \eg \cite{SP01}\,2001;
  \cite{MA05}\,2005, hereafter MAMB05) and the early light curves of
  GRBs (\cite{SP95}\,1995, 1997; \cite{DM98}\,1998). One dimensional
  (\cite{KI04}\,2004; MAMB05) and two dimensional (\cite{MA04}\,2004,
  hereafter MAMB04) simulations of internal shocks in relativistic
  outflows performed recently show that their evolution is
  considerably more complex than what can be inferred from approximate
  analytic models: the non-linear interaction of two shells leads to a
  merged shell which is very inhomogeneous. Hence, simple {\it
    one-zone} models are of limited validity when trying to infer
  physical properties of the emitting region from the flares
  (resulting from the collisions of shells). In MAMB05 we introduced a
  procedure to analyze spacetime properties of the emitting regions in
  relation to the shape of a flare. We showed that under certain
  conditions one can extract flow parameters not directly accessible
  by current observations, like the ratio of the Lorentz factors of
  the forward and reverse shocks (resulting from the collision of the
  shells), and the shell crossing times of these shocks.
  
  In the present work we extend our simulations to relativistic
  magnetohydrodynamic (RMHD) flows. To this end we have developed a
  RMHD version of the code presented in MAMB04 and MAMB05 which we
  call MRGENESIS. Using this new code we performed a systematic study
  of the influence of the initial shell magnetization on the internal
  shock dynamics and on the emitted radiation. Since the new code
  enables us to treat dynamic magnetic fields, we can use it to
  quantify the accuracy of the assumptions made in previous works
  about the constancy and homogeneity of the ratio of the magnetic
  field energy density and the internal energy density in relativistic
  outflows with internal shocks (see \eg \cite{DM98} 1998; MAMB04).
  
  In Sect.\,\ref{initial} we discuss in detail the initial properties
  of the colliding shells and describe the numerical method used to
  simulate their evolution. The interaction of two colliding shells
  and of the shells with the external medium are discussed in
  Sect.\,\ref{shells_medium}. The results of our numerical simulations
  are described in Sect.\,\ref{numerical}. We discuss and
  summarize our main findings in Sect.\,\ref{discussion}.

\section{Initial shell properties} 
  \label{initial}
  
  A probable cause of the observed blazar flares is the interaction of
  blobs (or shells) of matter within a relativistic outflow (jet),
  propagating at slightly different velocities. Such an interaction
  happens every time two shells collide after some time depending on
  their relative velocity. Internal shocks produced by the collision
  cause an enhanced emission of radiation which is thought to give
  rise to the observed flares.
  
  The interaction of the shells is simulated using a relativistic
  magnetohydrodynamic (RMHD) numerical scheme (\cite{LA05} 2005)
  coupled to the non-thermal radiation scheme developed in MAMB04. The
  resulting code MRGENESIS allows us to simulate the dynamic evolution
  of the magnetic field in a plasma instead of assuming that the field
  is randomly oriented in space and that its energy density is
  proportional to the internal energy density of the plasma, as was
  the case in our previous investigation (MAMB05). Thereby we are able
  to check whether the latter assumption, which is widely adopted in the
  literature, actually holds in the course of prototypical two-shell
  interactions.

  In order to simplify further discussion we introduce two
  magnetization parameters. The ratio of the magnetic energy density
  and the internal energy density of the plasma
  \footnote{Note that there is a typo in the equation defining $\ab$ in
  MAMB04. The left hand side of this equation should read $B^2$ with
  $B$ denoting the magnetic field strength in the comoving frame.}
  \begin{equation}\label{eq:alphab}
    \ab := \dsfrac{\dsfrac{1}{4\pi}\left(\dsfrac{B^2}{\Gamma^2}
           +\left({\mathbf v} \cdot {\mathbf B}\right)^2\right)}{ 
           \dsfrac{p}{\gad-1}} = (\gad-1)\dsfrac{2\pmag}{p}\, ,
  \end{equation}
  and the ratio of the magnetic energy density and the mass-energy
  density
  \begin{equation}\label{eq:sigma}
    \sigma := \dsfrac{1}{4\pi\rho}\left(\dsfrac{B^2}{\Gamma^2}
              +\left({\mathbf v} \cdot {\mathbf B}\right)^2\right) 
            = \dsfrac{2\pmag}{\rho}\, ,
  \end{equation}
  where $\Gamma$ is the bulk Lorentz factor of the flow, and where
 $\gad$ and $p$ are the adiabatic index and the thermal pressure of
 the plasma, respectively.
  \begin{equation}\label{eq:pmag}
    \pmag := \dsfrac{1}{8\pi}\left(\dsfrac{B^2}{\Gamma^2}
             +\left({\mathbf v}\cdot{\mathbf B}\right)^2\right)\,
  \end{equation}
  is the magnetic pressure in the comoving frame of a fluid element
  moving with a velocity $\mathbf v$ (and a corresponding Lorentz
  factor $\Gamma$). The magnetic field $\mathbf B$ is measured in the
  laboratory (source) frame, and we have chosen (here and in the
  following sections) units such that the speed of light $c = 1$, and
  that the strength of the magnetic field is measured in Gauss.

  We restrict ourselves to the simulation of one dimensional models,
  because we have shown previously that the lateral expansion is
  negligible during the collision of aligned shells, \ie all essential
  features can be captured using 1D simulations (MAMB04). We assume
  that the magnetic field is oriented perpendicular to the direction
  of the flow. Hence, $\pmag = B^2 / (8\pi\Gamma^2)$ and $\sigma = B^2
  / (4\pi\rho\Gamma^2)$. In this special case the comoving magnetic
  field strength $\BP$ is given by the field strength in the source
  frame through the relation $\BP = B / \Gamma$.  Initially the
  comoving magnetic field everywhere has the value $\BP = B / \Gamma =
  \sqrt{ 4\pi / (\gamma_{\rm ad}-1)\, \ab p} = 0.61 \sqrt{ (\ab^0 /
  10^{-2})\; p / (1.1\times 10^{-4}\, \rho_{\rm ext})}\,$mG. In the
  latter expression $p$ is assumed to be given in units of $\rho_{\rm
  ext} c^2$, and $\rho_{\rm ext}$ in turn in units of
  $10^{-23}\,$g/cm$^3$, respectively.
   
  In all relativistic hydrodynamic (RHD) models simulated in our
  previous studies (MAMB04, MAMB05) the magnetization parameters were
  (initially) $\ab = 10^{-3}$ and $\sigma = 10^{-7}$, respectively.
  For the RMHD investigation presented here we have evolved a set of
  models whose (hydrodynamic) parameters are similar to those of model
  \SF{10}{10} of MAMB05. In this (hydrodynamic) model two shells of
  uniform rest mass density $10^{-20}\,$g\,cm$^{-3}$, initial
  thickness and separation $10^{14}\,$cm propagate through a
  homogeneous ambient medium (Lorentz factor $\Gamma_{\rm ext} = 2.9$,
  density
%\footnote{\bf This normalization of the ambient medium
%    density is chosen to allow the magnetic field in the shells to
%    remain in the observationaly constrained value of $\simgt 0.01 -
%    1$ G during the time of emission.}  
  $\rho_{\rm ext} = 10^{-23}\,$g\,cm$^{-3}$) with Lorentz factors
  $\Gamma_1 = 5$ and $\Gamma_2 = 7$, respectively. The emitted
  radiation is computed using the type-E shock acceleration model with
  $\alpha_{\rm e} = 10^{-2}$, $\gamma_{\rm min} = 30$, and $\eta = 7
  \times 10^3$ (see MAMB04 for details).  We simulated two groups of
  RMHD models (see Table\,1):
  \begin{enumerate}
  \item The parameters of models \BG{1} to \BG{6} are identical to
    those of model \SF{10}{10} of MAMB05 except for the magnetization
    parameter $\ab^0$ which varies from $10^{-1}$ to $10^{-4}$.  This
    subset of models allows us to study the influence of the initial
    magnetization on the light curve and its impact on the dynamics.
  \item In models \TG{1} to \TG{4} the strength of the magnetic field
    is kept constant and equal to that of model \BG{4}, but the value
    of the initial thermal pressure of the plasma is varied within a
    factor of $\sim 20$.
  \end{enumerate}

% ... 1D simulations table
  \begin{table}
    \begin{center}
      \begin{tabular}{|c||c|c|c|c|c|}
        \hline
          {\tiny model} & $\ab^0$ & $\sigma^0$ & $(p/\rho)^0$ 
                & $\BP_{\rm shell}\, [{\rm mG}]$ 
                & $\BP_{\rm   ext}\, [{\rm mG}]$ 
        \\ \hline\hline
  \BG{1} & $        10^{-1}$ & $3.3\times 10^{-5}$ & $1.1\times 10^{-4}$ & $61.4$ & $1.94$ \\ 
  \BG{2} & $8\times 10^{-2}$ & $2.6\times 10^{-5}$ & $1.1\times 10^{-4}$ & $54.9$ & $1.74$ \\  
  \BG{3} & $5\times 10^{-2}$ & $1.7\times 10^{-5}$ & $1.1\times 10^{-4}$ & $43.4$ & $1.37$ \\   
  \BG{4} & $        10^{-2}$ & $3.3\times 10^{-6}$ & $1.1\times 10^{-4}$ & $19.4$ & $0.61$ \\  
  \BG{5} & $        10^{-3}$ & $3.3\times 10^{-7}$ & $1.1\times 10^{-4}$ & $ 6.1$ & $0.19$ \\  
  \BG{6} & $        10^{-4}$ & $3.3\times 10^{-8}$ & $1.1\times 10^{-4}$ & $ 1.9$ & $0.06$ \\  
  \TG{1} & $7\times 10^{-2}$ & $3.3\times 10^{-6}$ & $1.6\times 10^{-5}$ & $19.4$ & $0.61$ \\  
  \TG{2} & $3\times 10^{-2}$ & $3.3\times 10^{-6}$ & $3.7\times 10^{-5}$ & $19.4$ & $0.61$ \\  
  \TG{3} & $7\times 10^{-3}$ & $3.3\times 10^{-6}$ & $1.6\times 10^{-4}$ & $19.4$ & $0.61$ \\  
  \TG{4} & $3\times 10^{-3}$ & $3.3\times 10^{-6}$ & $3.7\times 10^{-4}$ & $19.4$ & $0.61$ \\  
        \hline
      \end{tabular}
    \end{center}
    \caption{Properties of the simulated models where two shells of
      uniform initial rest mass density $\rho_1 = \rho_2 =
      10^{-20}\,$g\,cm$^{-3}$ propagate through a homogeneous ambient
      medium (Lorentz factor $\Gamma_{\rm ext} = 2.9$, density
      $\rho_{\rm ext} = 10^{-3}\times \rho_{\rm shell} =
      10^{-23}\,$g\,cm$^{-3}$) with Lorentz factors $\Gamma_1 = 5$ and
      $\Gamma_2 = 7$, respectively.
      $\ab^0$ and $\sigma^0$ are the initial uniform magnetization
      parameters (ratio of magnetic energy density and internal energy
      density, and ratio of magnetic energy density and mass-energy
      density, respectively), $(p/\rho)^0$ is the initial uniform
      thermal pressure of the plasma in units of its rest mass energy
      density, and $\BP_{\rm shell}$ and $\BP_{\rm ext}$ give the
      initial value of the comoving magnetic fields of the shells and
      of the external medium, respectively. Note that the magnetic
      field is oriented perpendicular to the flow direction, and that
      the initial conditions of model \BG{5} are identical to those of
      model \SF{10}{10} of MAMB05.}
    \label{tab:1}
  \end{table}

  The relative Lorentz factor between the shells $\Gamma_{\rm
    rel}=\Gamma_1\Gamma_2\left(1-\sqrt{(1-\Gamma_1^{-2})(1-\Gamma_2^{-2})}\right)$
  fulfills the condition $\Gamma_{\rm rel}>\sqrt{1+\sigma}$ under
  which of a pair of internal shocks forms (e.g. \cite{FA04} 2004;
  \cite{ZK05} 2005).

  Although our models are computed in one spatial dimension, our
  simulations are effectively two dimensional. The second dimension is
  spanned by the energy of the electrons. It is discretized into 64
  energy bins. As mentioned above, we have assumed a ratio of shell to
  ambient density $\chi_\rho=10^3$. This value is a compromise between
  those numerically feasible and those expected in blazars. Larger
  density ratios fit the blazar values better, but cause very small
  time steps. For our value each model requires about two weeks of
  computing time on an IBM Power IV processor, while a model with
  $\chi_\rho=10^5$ needs about one month of computing time. We point
  out that the value we have chosen is sufficiently large to account
  for all qualitative behaviours expected for even larger values of
  $\chi_\rho$.

\section{Interaction of shells with the external medium}
  \label{shells_medium}

  Since the shells move faster than the ambient medium, their front
  edges interact with the ambient plasma. The initially sharp edges of
  the shells evolve into a set of simple waves whose position and
  physical characteristics are calculated by means of an exact RMHD
  Riemann solver (\cite{RO05} 2005). For this calculation we assume
  that the evolution of the shell edges is one dimensional (\ie along
  the direction of motion), and that the magnetic field is oriented
  perpendicular to the velocity of the shells. According to this
  set-up, we model blobs hat experience a negligible transverse
  expansion when propagating away from the galactic nucleus, and which
  carry along magnetic fields preferentially oriented perpendicular to
  the blob direction of motion. A similar situation is encountered
  when modeling magnetized shells emerging from the progenitor system
  of a GRB except that the Lorentz factor of the shells is $\sim
  100$.

\begin{figure}
    \centering \includegraphics[scale=0.3]{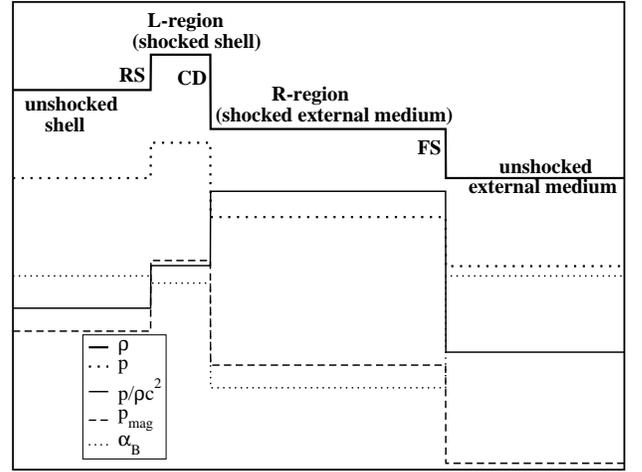}%

    \caption{Flow structure near the front part of a
      relativistic shell moving through an ambient medium (to the
      right). The initial density and pressure of the shell is $1000$
      times larger than that of the ambient medium. The shell
      propagates with a Lorentz factor $\Gamma = 17$, and the initial
      ratio of pressure and rest-mass energy density is $10^{-4}$. Two
      shocks form, a reverse shock (RS) propagating into the shell,
      and a forward shock (FS) propagating into the ambient
      medium. The shocked shell and the ambient medium are separated
      by a contact discontinuity (CD). The various lines show (from
      top to bottom) the rest mass density (thick solid line), the
      thermal pressure (thick dotted line), the ratio of thermal
      pressure and rest mass energy (thin solid line), the magnetic
      pressure (thin dashed line), and the ratio of magnetic energy
      density and thermal energy density (thin dotted line) using
      arbitrary units. }
\label{fig:flow}
\end{figure} 

  A prototypical evolution of the front edge of a shell is shown in
  Fig.\,\ref{fig:flow}. In the particular case displayed here, the
  external medium is assumed to be at rest, while the shell moves with
  $\Gamma = 17$. Both its density and pressure are $1000$ times larger
  than in the ambient medium. Inside the shell as well as in the
  ambient medium the magnetic field strength is such that $\ab =
  10^{-3}$. The thermal pressure is $1.11 \times 10^{-3} \rho$
  everywhere. Under these conditions two shocks form, a forward and a
  reverse one, which are separated by a contact discontinuity, and
  which propagate into the ambient medium and into the shell,
  respectively. The structure is qualitatively similar to the one
  found in the pure RHD case (see \eg Fig.\,2 of MAMB04). However, the
  shocks are not purely hydrodynamic ones, but fast hydromagnetic
  shocks.
\footnote{Note that as the magnetic field is assumed to be oriented
  perpendicular to the flow direction, fast and slow magnetosonic waves
  as well as Alfv\'en waves propagate with the same velocity.}
  Thus, they amplify the magnetic field due to the compression of the
  plasma: the front shock compresses the magnetic field of the ambient
  medium (if non-zero), while the reverse shock amplifies the shell's
  magnetic field.
  
  The ratio of the magnetic energy density and the thermal energy
  density $\ab$ (thin dotted line in Fig.\,\ref{fig:flow}) decreases
  in the shocked regions. The magnetization parameter $\sigma$
  \emph{increases} in the shocked shell since $B/(\rho\Gamma)$ is
  uniform across shocks and rarefaction when the magnetic field is
  perpendicular to the velocity (\cite{RO05} 2005), \ie $\sigma
  \propto B^2/\rho \propto \rho$. The particular case shown in
  Fig.\,\ref{fig:flow} represents the qualitative evolution of a
  single shell. However, as the quantitative properties of the shocked
  region depend on the Lorentz factor of the shell, we have performed
  a parameter study to determine these. The four panels of
  Fig.\,\ref{fig:variations} show the variation of the density, the
  pressure, and the magnetization parameter $\ab$ in the shocked
  regions (measured relative to the corresponding quantities in the
  unshocked regions) with the shell Lorentz factor.  Several
  remarkable points can be inferred from Fig.\,\ref{fig:variations}:
  \begin{figure*}
    \centering
    \includegraphics[scale=0.3]{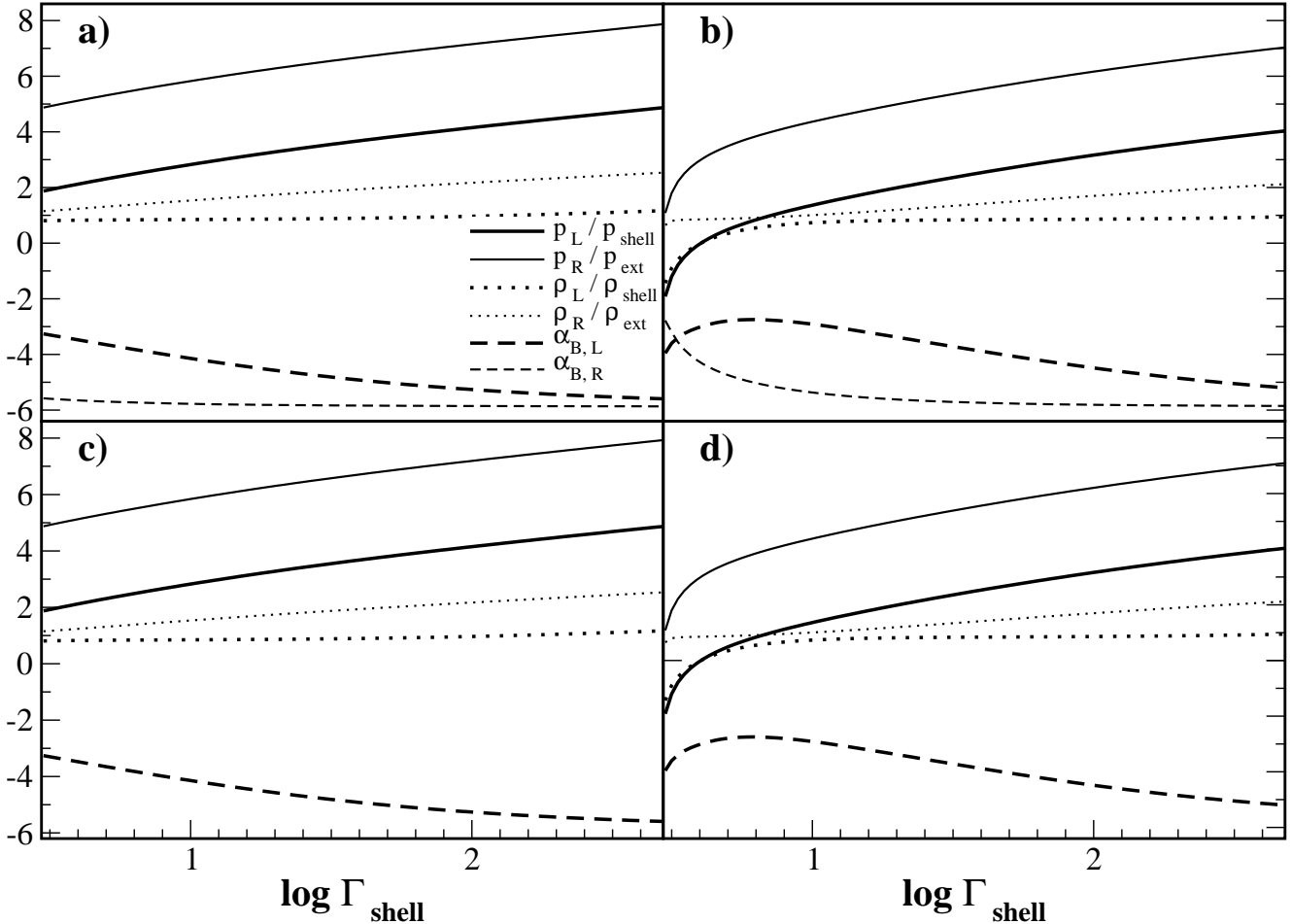}
    \caption{Variation of the logarithms of $\rho_L / \rho_{\rm
      shell}$, $\rho_R / \rho_{\rm ext}$, $p_L / p_{\rm shell}$, $p_R
      / p_{\rm ext}$, $\alpha_{\rm B,L}$, and $\alpha_{\rm B,R}$ as a
      function of the shell Lorentz factor, $\Gamma_{\rm shell}$.
      Subscripts $L$ and $R$ denote the shocked regions to the left
      and right of the contact discontinuity, respectively. The
      density and the pressure of the shell are $1000$ times larger
      than in the ambient medium, and $\alpha_{\rm shell} = 10^{-3}$
      (as in the case shown in Fig.\,\ref{fig:flow}). The parameters
      of the ambient medium are: (a) $(\Gamma_{\rm ext}, \alpha_{\rm
      B, ext}) = (1,10^{-3})$, (b) $(\Gamma_{\rm ext}, \alpha_{\rm B,
      ext}) = (3,10^{-3})$, (c) $(\Gamma_{\rm ext}, \alpha_{\rm B,
      ext}) = (1,0)$, and (d) $(\Gamma_{\rm ext}, \alpha_{\rm B, ext})
      = (3,0)$, respectively.}
    \label{fig:variations} 
  \end{figure*}

  \begin{enumerate}
        
  \item The ratio of the thermal pressure in the shocked shell region
    and the initial pressure rises with the Lorentz factor. Even for
    relatively small Lorentz factors ($\simgt 5$) the thermal pressure
    is $100$ times larger than the initial one in the case of an
    ambient medium at rest. The pressure of the ambient medium rises
    $100$ to $10^4$ times above its initial value. In the case of a
    moving ambient medium and for shell Lorentz factors $\simlt 4$ a
    rarefaction instead of the reverse shock forms ($\rho_{\rm L} /
    \rho_{\rm shell} < 1$; flow structure $\leftarrow \cal{R} \cal{C}
    \cal{S} \rightarrow$\footnote{In this notation $\cal{R}$
      represents a rarefaction, $\cal{C}$ a contact discontinuity and
      $\cal{S}$ a shock. The arrows indicate the direction of
      propagation of the rarefaction or of the shock with respect to
      the contact discontinuity.}), while for larger Lorentz factors
    the situation is as shown in Fig.\,\ref{fig:flow} ($\leftarrow
    \cal{S} \cal{C} \cal{S}\rightarrow$).
    
  \item A comparison of panels (a) and (c), or (b) and (d) shows that
    the properties of the shocked shell region only weakly depend
    on the magnetization of the ambient medium provided the latter is
    initially sufficiently small ($\alpha_{\rm B, ext} \leq 10^{-3}$)
    and the shell has $\Gamma_{\rm shell} \simgt 5$, \ie an initially
    weakly magnetized ambient medium has no influence on the evolution
    of the shell.

  \item The rest mass density in the shocked part of the shell ({\it
    L-region} in Fig.\,\ref{fig:flow}) is independent of the Lorentz
    factor for $\Gamma \simgt 5$. The comoving magnetic field $\BP =
    B/\Gamma$ is independent of the shell Lorentz factor, too, since
    $\BP / \rho = constant$ across shocks for magnetic fields
    perpendicular to the velocity field. Since $\ab \propto \rho^2/p$,
    $\rho_{\rm L} \simeq constant$, and $p_{\rm L}$ is monotonically
    increasing, $\alpha_{\rm B, L}$ decreases monotonically in the
    L-region.
    
  \end{enumerate}
  
  The latter fact must be taken into account in the internal shock
  scenario to properly estimate the strength of the magnetic field
  of the faster shell at the moment when the interaction starts (\ie
  when its front edge starts colliding with the back edge of the
  slower shell), since the magnetic field will be \emph{larger}, and
  $\ab$ will be \emph{smaller} than their corresponding initial
  values. 

  If a value $\chi_\rho > 10^3$ were used, the leading edges of both
  shells would develop a Riemann fan with the same qualitative
  structure. However, relative to the shell flow the two shocks
  emerging from the leading edges will propagate much more slowly, and
  accordingly the width of the Riemann fan will be smaller, too. In
  the limit $\chi_\rho\rightarrow \infty$ the width of the fan tends
  to zero, and the shells behave as rigid {\it bodies} propagating
  through the vacuum (which is the typical assumption made in analytic
  models of colliding shells).
  
  \cite{ZK05} (2005) have considered in detail the dependence of
  various flow parameters on the relative Lorentz factor between the
  shell and the ambient medium, and on the magnetization parameter
  $\sigma$. They also find that for a sufficiently large Lorentz
  factor, appropriately normalized flow parameters are insensitive to
  the Lorentz factor, depending only on the magnetization. For
  example, the dependence of the ratio of magnetic pressure and
  thermal pressure in the shocked region of the shell (see panel (e)
  in Fig.\,1 of their paper) is consistent with the trend exhibited in
  Fig.\,\ref{fig:variations}

\section{Numerical simulations}
  \label{numerical}

  \begin{figure}
    \centering \includegraphics[scale=0.3]{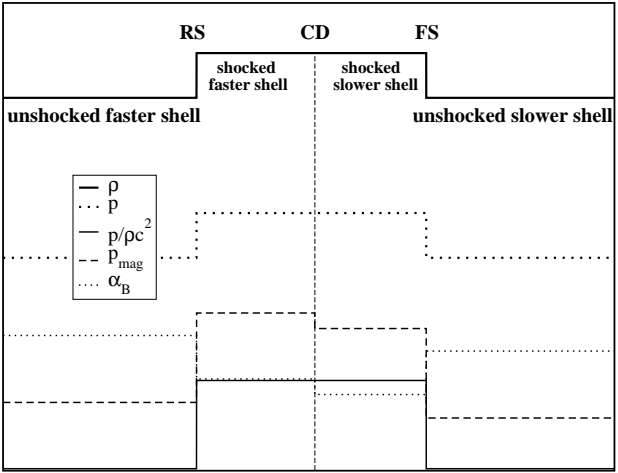}%

    \caption{Same as Fig.\, \ref{fig:flow}, but for the flow structure
    near the interface of two interacting shells (assuming no
    evolution prior to interaction).  The initial density, pressure
    and comoving magnetic field are assumed to be the same in both
    shells. The Lorentz factor of the faster (left) shell is $7$, and that
    of the slower (right) shell is $5$. }
    \label{fig:idealint}
  \end{figure}

  As in MAMB05, we simulate shell collisions in one spatial dimension
  only. This approach is justified as all essential features of the
  collision of aligned shells can be captured in one dimension, since
  the lateral expansion is negligible during the collision.

  The computational grid consists of $10^4$ equidistant zones covering
  a physical domain of length $5 \times 10^{15}\,$cm. A grid
  re-mapping technique (MAMB04) is applied in order to follow the
  evolution of two shells initially separated by $10^{14}$ cm up to
  distances of $10^{17}$ cm with sufficiently good resolution.
  
  The energy distribution of the non-thermal electrons is sampled by
  64 logarithmically spaced energy bins, and the synchrotron radiation
  is computed at 40 frequency values logarithmically spanning the
  frequency range from $10^{15}\,$Hz to $10^{20}\,$Hz. We use the
  type-E shock acceleration model of MAMB04, with some modifications
  as explained below.

  The RMHD conservation laws are integrated and the synchrotron
  radiation is computed using \emph{MRGENESIS}, which is an extension
  of the RMHD version of the RHD code \emph{GENESIS} (\cite{LA05}
  2005). For the adiabatic index of the plasma a value $\gamma_{\rm
  ad} = 4/3$ is assumed. The handling of the non-thermal particles and
  the computation of the non-thermal emission follows the procedure
  described in MAMB04, except for the following two modifications:

  \begin{figure*}
    \centering
      \includegraphics[scale=0.3]{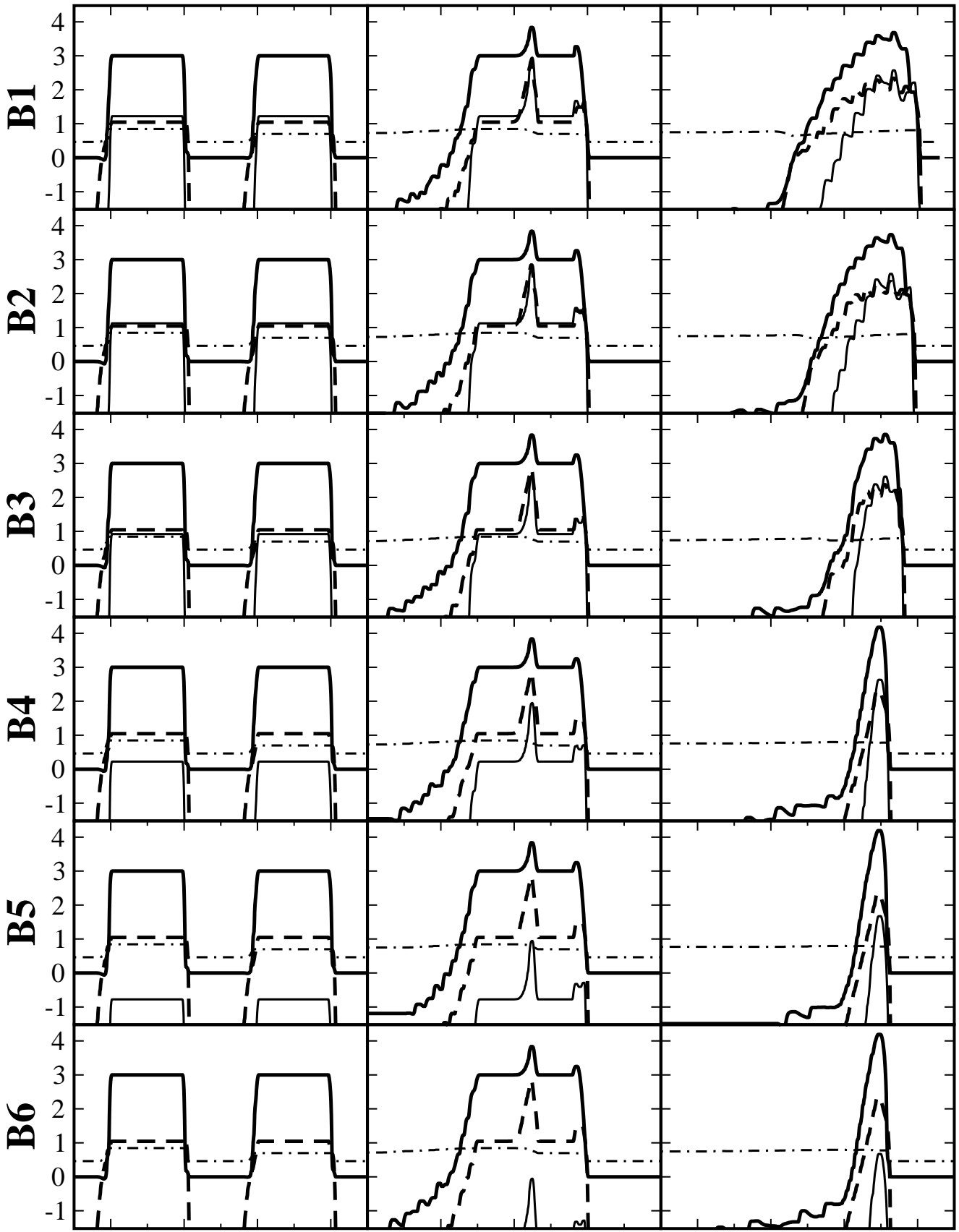}
      \caption{ Three snapshots from the evolution of models
        \BG{1}-\BG{6}. The thick solid line shows the logarithm of the
        rest-mass density, the dashed line the thermal pressure, the
        thin solid line the comoving magnetic field energy density,
        and the dash-dotted line the Lorentz factor, respectively. The
        panels in the first column display the situation $1.5\,$ks
        after the start of the simulations, the second column after
        $500\,$ks, and the third column after $2000\,$ks (all times
        are measured in the source frame). The size of the horizontal
        axis corresponds to a length of $4\times 10^{14}\,$cm. }
    \label{fig:mhdmodels} 
  \end{figure*}

  \begin{figure*}
    \centering
      \includegraphics[scale=0.3]{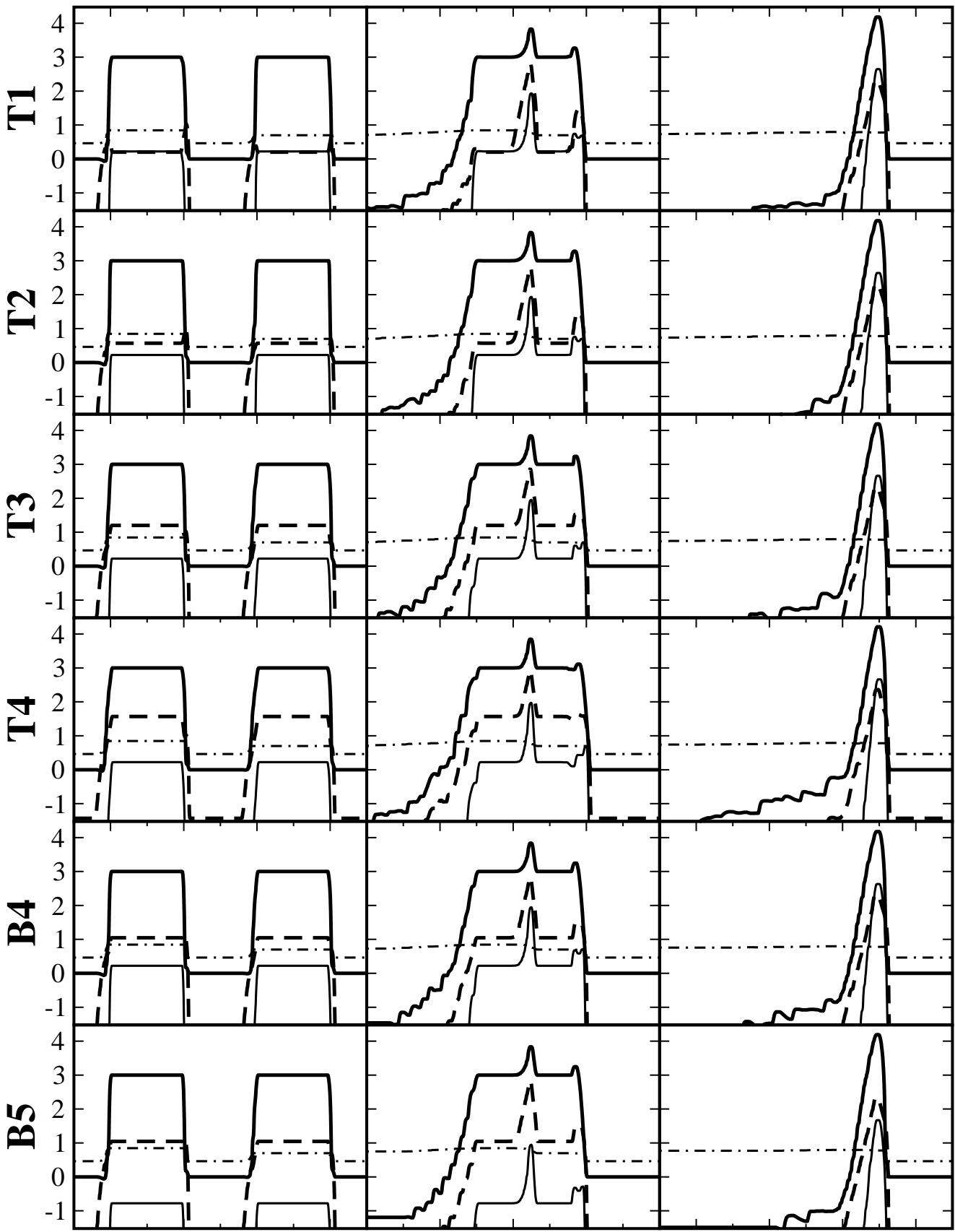}
      \caption{Same as Fig.\,\ref{fig:mhdmodels}, but for models
               \TG{1}-\TG{4}. The snapshots from models \BG{4} and
               \BG{5} are shown here again to allow for an easier
               comparison between the models of the {\bf T} and {\bf
               B} series.}
   \label{fig:mhdmodels1} 
    \end{figure*}

  \begin{enumerate}
    
  \item {\bf Synchrotron radiation}: for each grid zone the
    dynamically evolving magnetic field ${\mathbf B}$ (measured in the
    source frame) is transformed into the comoving frame defined by
    the flow velocity ${\mathbf v}$ (or Lorentz factor $\Gamma$) of
    the zone in order to compute the synchrotron emissivity according
    to Eq.\,14 of MAMB04. However, instead of the magnetic field in
    the comoving frame $\BP$ we insert $\BP \sin \theta^{\prime}$,
    where $\theta^{\prime}$ is the angle (in the comoving frame)
    between the line of sight towards the observer and the magnetic
    field vector. In the present work the flow is assumed to be
      propagating towards an observer and parallel to the line of
       sight. Since also the magnetic field is assumed to be initially
      perpendicular to the direction of the flow, we get $\BP = B /
    \Gamma$ and $\theta^{\prime} = \pi/2$.

  \item {\bf Acceleration of non-thermal particles}: we assume that
    non-thermal particles are accelerated in the shocks that are
    detected in the magnetized flow with the standard PPM routines
    implemented in \emph{GENESIS} (\cite{AL99} 1999). However, we
    replace the jumps in gas pressure by the jumps in total pressure
    ($p^* \equiv p+\pmag$) in the shock detection algorithm.

  \end{enumerate}

  \subsection{Magnetohydrodynamic evolution}
    \label{evolution:MHD}
    
    Figure\,\ref{fig:idealint} shows an example of the analytic
    solution of the interaction of two shells, assuming they have
    undergone no evolution prior to their collision. This is an
    approximation of the more realistic situation where the
    pre-collision evolution changes the conditions at the edges of the
    shells that are interacting, which is the case of the detailed
    numerical simulations presented here and in MAMB05. Nonetheless,
    this approximation captures the qualitative structure of the
    Riemann fan emerging from interacting over-dense shells and,
    particularly the fact that the density jump at the contact
    discontinuity is expected to be small (which can also be seen in
    the results of our numerical simulations;
    Figs.\,\ref{fig:mhdmodels} and \ref{fig:mhdmodels1}, where the
    evolution prior to collision was taken into account).

    Figures \ref{fig:mhdmodels} and \ref{fig:mhdmodels1} show three
    snapshots of the magnetohydrodynamic evolution of the $10$ models
    considered in this work. The left column displays the initial
    state, the middle column the propagation of the internal shocks
    through the shells after $500$\,ks, and the right column the break
    out of the shocks from the merged shell after $2000$\,ks.
    
    In the middle column the two internal shocks flanking the dominant
    peak inside the shells can be noticed (corresponding to the region
    between RS and FS in Fig.\,\ref{fig:idealint}) as well as the
    front structure of the slower (right) shell, which arises from the
    interaction of the shell with the ambient medium (see also
    Fig.\,\ref{fig:flow}). For models \TG{1} - \TG{4} the thermal
    pressure within the front structure is different, \ie larger
    values of $\ab^0$ (for a fixed magnetic field; see Table\,1) yield
    a greater thermal pressure contrast between the slow shell and the
    shocked ambient medium ahead of it.  Analytically, it is possible
    to show that the reverse shock propagates more slowly through the
    merged shell as $\ab^0$ decreases.  This trend is exhibited by the
    models of the \BG{}--series (Fig.\,\ref{fig:mhdmodels}), where
    $\ab$ decreases from $10^{-1}$ (model \BG{1}) to $10^{-4}$ (model
    \BG{6}; see Table\,1).

    For the models of the T-series the differences in the morphology
    and the dynamics of the merged shell are not as pronounced as for
    the B-models, because the initial magnetization of the former
    models differs only by a fraction $\ab^0(T1) / \ab^0(T4) \approx
    20$, while $\ab^0(B1) / \ab^0(B4) \approx 10^3$.  However, the
    slowest shell of model \TG{4}, which has the largest thermal
    pressure, develops a reverse rarefaction instead of a shock as a
    result of its interaction with the external medium (see middle
    panel of Fig.~\ref{fig:mhdmodels1}), although this difference does
    not have any noticeable influence on the long term evolution of
    the merged shell.
    More relevant in the T-series is the near independence of the
    results on the initial ratio $(p/\rho)^0$. Hence, if the initial
    shells are sufficiently ``cold'' (\ie $(p/\rho)^0 \simlt 10^{-3}$)
    the exact value of $p$ does not influence the evolution. This
    implies that it may be impossible to infer the (absolute) value of
    the initial shell pressure from observations.
    
    The final thickness of the merged shells (after 2000\,ks) depends
    monotonically on $\ab^0$, because the total pressure
    $p_\mathrm{tot} = p\, \{1 + \ab / [ 2 (\gad - 1) ] \}$ increases
    with $\ab^0$. Model \BG{1}, which has the largest initial
    magnetization of all models ($\ab^0 = 0.1$), gives rise to the
    thickest merged shell with $\sim 2 \times 10^{14}\,$cm, which is
    of the order of the sum of the initial sizes of the colliding
    shells. Model \BG{6} produces the thinnest merged shell with $\sim
    7\times 10^{13}\,$cm, which is less than the initial size of a
    single shell. Hence, the final size of the merged shells differs
    little from the sum of the sizes of the two initial shells. This
    can be understood from the fact that the final size of the merged
    shells results from the competition of two factors: the expansion
    triggered when the colliding shells are heated up by internal
    shocks, and the compression produced by the collision.

  \subsection{Energy conversion efficiency}
    \label{evolution:efficiency}

    In MAMB05 we monitored the efficiency of the conversion of bulk
    kinetic energy into internal energy during the collision of the
    shells. The knowledge of the efficiency is relevant to make
    predictions about the amount of radiated energy in any model
    invoking internal shocks as the primary source of the non-thermal
    emission of an astrophysical plasma. Here we extend our previous
    study to magnetized shells, \ie the bulk kinetic energy of the
    shells can be converted either into thermal energy or into
    magnetic energy. 

    We fist define the total instantaneous efficiency of converting
    bulk kinetic energy into magnetic and thermal energy as
% eps
    \begin{equation}\label{eq:eps}
      \epsilon(T) := 1 - \dsfrac{\int \Gamma(z, T)(\Gamma(z, T) - 1)\ \rho(z, T)\ \difd z}
	      {\int \Gamma(z, 0)(\Gamma(z, 0) - 1)\ \rho(z, 0)\ \difd z}\, .
    \end{equation}
% end of eps

\begin{figure}
      \centering
      \includegraphics[scale=0.32]{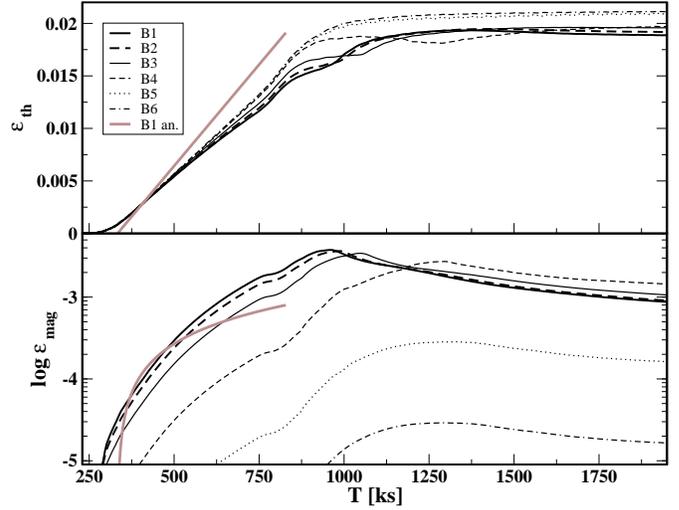}
      \caption{Efficiency of converting bulk kinetic energy into
        thermal (upper panel) and magnetic field energy (lower panel)
        for models of the \BG{}-series. Times are measured in the
        source frame. The thick grey line gives an analytic
        approximation (see Appendix \ref{app:eff} for details).}
      \label{fig:eff1}
    \end{figure}

\begin{figure}
      \centering
      \includegraphics[scale=0.32]{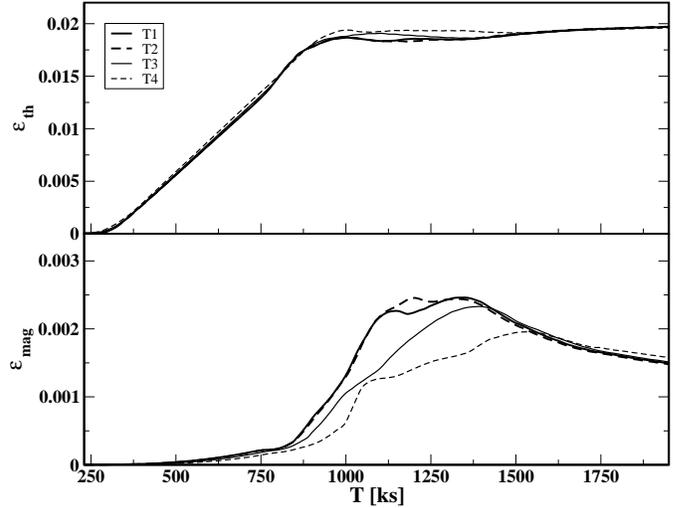}
      \caption{Same as Fig.\,\ref{fig:eff1}, but for models of the
        \TG{}-series. The graphs for models \BG{4} and \BG{5} are
        added for comparison.}
      \label{fig:eff2} 
    \end{figure}

    Next, using the standard relativistic magnetohydrodynamic
    expression for the total energy density measured in the laboratory
    frame $\tau := \rho {\tilde h} \Gamma^2 - p - p_{\rm mag} - \rho
    \Gamma$ (where ${\tilde h} := 1 + \gad p/ ((\gad-1)\rho) + \sigma$
    is the total enthalpy per unit of mass including the magnetic
    contribution), we can identity the terms corresponding to the
    kinetic, the thermal and the magnetic energy densities.  Taking
    into account that initially there is not only kinetic energy in
    the flow but also thermal and magnetic energy, we define the
    efficiency of converting kinetic energy into thermal energy as
    \begin{equation}\label{eq:eps_thp}
      \begin{array}{rcl}
        \epsth'(T) & :=& \biggl[ \displaystyle 
                        \int\ \left(\dsfrac{\gad}{\gad-1}\Gamma^2(z,T) 
                                    - 1\right)\  p(z,T)\ \difd z \\
                  &   & - \displaystyle 
                        \int\ \left(\dsfrac{\gad}{\gad-1}\Gamma^2(z,0) 
                                    - 1\right)\  p(z,0)\  \difd z\biggr] \\
                  &   & \displaystyle / \displaystyle 
                        \int\ \Gamma(0,T)(\Gamma(0,T)-1)\ 
                               \rho(0,T)\ \difd z
      \end{array} \, ,
    \end{equation}
    and the efficiency of converting kinetic energy into magnetic
    field energy as
    \begin{equation}\label{eq:eps_magp}
      \begin{array}{rcl}
        \epsmag'(T) &:=& \biggl[\displaystyle 
                        \int\ \left(1-\dsfrac{1}{2 \Gamma^2(z,T)}\right) 
                        \dsfrac{B^2(z, T)}{4\pi}\ \difd z \\
                   &  & - \displaystyle 
                        \int\ \left(1-\dsfrac{1}{2 \Gamma^2(z,0)}\right) 
                        \dsfrac{B^2(z,0)}{4\pi}\  \difd z\biggr] \\
                   &  & \displaystyle /
                        \int\ \Gamma(0,T)(\Gamma(0,T)-1)\ \rho(0,T)\ \difd z
      \end{array} \, ,
    \end{equation}
    respectively.  The quantities $\Gamma$, $p$, $\rho$, and $B$ that
    appear in the integrals are measured either at time $t=T$ or at
    time $t=0$, and at position $z$. The integrals in
    Eqs.\,\ref{eq:eps_thp} and \ref{eq:eps_magp} should extend over
    the whole domain swept up by the shells until the end of the
    simulation, but we restrict the domain of evaluation to that of
    the instantaneous computational grid. This restriction is
    justified since the dominant contribution to the energy (either
    kinetic, thermal or magnetic) comes from the shells and the
    shocked ambient medium, which are both always covered by the
    computational grid, regardless of the grid-remapping (see
    above). The contributions ignored due to grid-remapping are thus
    negligible regarding the dynamic evolution of the system, but may
    effect the efficiencies as they are of the same order of magnitude
    ($\simlt 1\%$). Therefore, we re-define the efficiencies given in
    Eqs.\ref{eq:eps_thp} and \ref{eq:eps_magp} by scaling them with
    $\epsilon(T)$, \ie
    \begin{equation}\label{eq:eps_th}
      \epsth(T) = \dsfrac{\epsth'(T)}{\epsth'(T) 
                + \epsmag'(T)}\epsilon(T) \, ,
    \end{equation}
    and
    \begin{equation}\label{eq:eps_mag}
      \epsmag(T) = \dsfrac{\epsmag '(T)}{\epsth'(T) 
                 + \epsmag'(T)}\epsilon(T)\, .
    \end{equation}
    Figs.\,\ref{fig:eff1} and \ref{fig:eff2} show the evolution of
    $\epsth$ (upper panel) and $\epsmag$ (lower panel) for models
    \BG{1} to \BG{6}, and models \TG{1} to \TG{4}, respectively. From
    the upper panel of Fig.\,\ref{fig:eff1} one infers that the
    efficiency decreases with $\ab$ at late times for all models.  The
    lower panel of Fig.\,\ref{fig:eff1} displaying the conversion of
    kinetic energy into magnetic field energy confirms that models
    with $\ab\leq 0.01$ behave qualitatively differently from models
    with $\ab>0.01$, as the global maximum of the efficiency of the
    latter models is shifted to earlier times than in models with
    $\ab\leq 0.01$. The upper panel of Fig.\,\ref{fig:eff2} shows that
    the thermal efficiencies $\epsth$ of models \TG{1} to \TG{4} are
    similar to that of model \BG{4}, and that the evolution is almost
    identical for all four models. The efficiency $\epsmag$ (lower
    panel) is generally larger for a lower initial thermal pressure
    (see Table,1).

    In Appendix \ref{app:eff} we give the details of an analytic
    approximation to compute the efficiencies. The approximation
    takes into account that the shells are compressible {\it fluids}
    and not {\it solids} that collide inelastically, as has been
    widely assumed in the literature. The thick grey line in
    Fig.\,\ref{fig:eff1} shows our analytic approximations for
    $\epsth$ (upper panel) and $\epsmag$ (lower panel),
    respectively. The parameters for the analytic model are the same
    as in model \BG{1}.

  \subsection{Light curves}
    \label{lc}

    For each model two light curves were computed, one in the energy
    range $0.1 - 1$\,keV (soft band), and second one in the energy
    range $2 - 10$\, keV (hard band). The bands are defined to
    approximately match those used by \cite{BR05} (2005) when
    analyzing their observations. As described in detail in MAMB04,
    light curves are computed by summing the contributions of the
    emissivity from each grid zone in each time step taking into
    account the light travel time to the observer.

    \begin{figure*}
      \centering
      \includegraphics[scale=0.43]{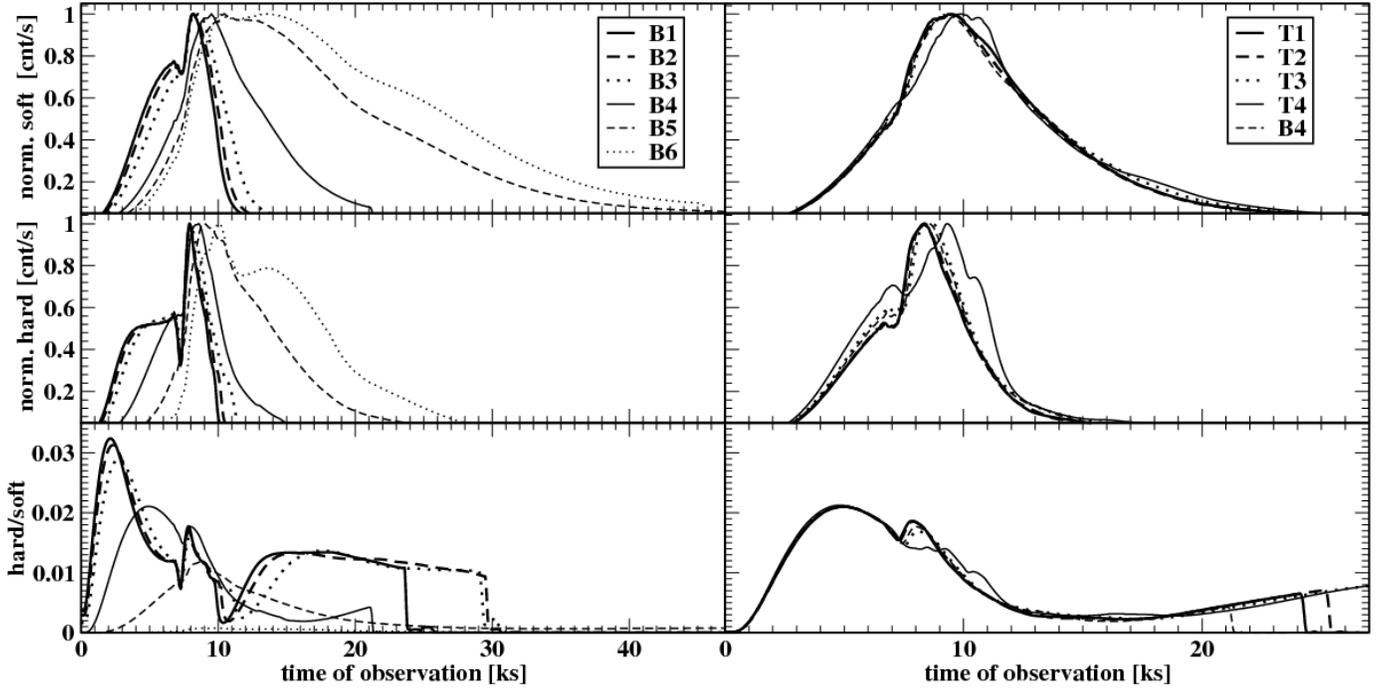}
      \caption{Normalized soft (top) and hard (middle) light curves,
        and instantaneous hardness ratio (bottom) for models \BG{1} -
        \BG{6} (left), and \TG{1} - \TG{4} and \BG{2} (right),
        respectively. Note that all light curves have been normalized
        individually. }
      \label{fig:LC} 
    \end{figure*}
    
    Figure\,\ref{fig:LC} shows the normalized soft (upper panels) and
    hard (middle panels) light curves of our $10$ models. The
    difference between the light curves of the \BG{}-models (left) and
    \TG{}-models (right) is obvious, as well as their similarity for
    all \TG{}-models and model \BG{4}. The instantaneous hardness
    ratio (HR), defined as the ratio of hard and soft counts at a
    given observational time (see bottom panels) rapidly rises during
    the collision phase, and then decreases with the observational
    time. Hence, at late times ($t \simgt 15\,$ks) the detected
    radiation becomes softer. Indeed, for models \BG{1} to \BG{3} the
    light curves and the hardness ratio have a multi-peaked structure,
    a behavior which has been observed for Mrk\,421 (\cite{BR05} 2005)
    or PKS2155-304 (\cite{ZA06} 2006). In these models having a high
    magnetization ($\ab > 5 \times 10^{-2}$) the relative amount of
    radiation emitted in both bands changes rapidly with time.  Models
    with $\ab > 10^{-2}$ show a persistent, moderately high hardness
    ratio ($>0.01$) at late epochs ($t>15\,$ks), which correlates with
    the fact that the RMHD evolution of the merged shell produced by
    initial shells with $\ab < 10^{-2}$ is characterized by a
    long-persisting structure radiating dominantly in the soft band
    (see Sect.\,\ref{concl:radiation}). \TG{}-models, all of them with
    low initial magnetic field (equal to that of model \BG{4}) do not
    exhibit a variable behavior but the same smooth trends as models
    \BG{5} and \BG{6}.

\section{Discussion}
  \label{discussion}

  In the following sections we will summarize the main results of our
  numerical investigation.

 \subsection{Magnetohydrodynamic  evolution}
   \label{concl:interaction} 

   MAMB05 found that the relativistic hydrodynamic (RHD) evolution of
   shell interactions can be divided into three phases: pre-collision,
   collision, and post-collision (rarefaction) phase. The evolution of
   magnetized shells also exhibits these three phases, but with some
   differences.

   \subsubsection{Pre-collision phase}

   Since the magnetic field inside the shell is perpendicular to the
   direction of motion, the initially sharp forward edge of the shell
   develops into a Riemann fan consisting of three simple waves,
   namely two magnetosonic shocks separated by a contact discontinuity
   (Fig.\,\ref{fig:flow}). The reverse shock compresses the shell as
   it propagates towards the shell's rear edge. The compression factor
   ($\rho_{\rm L} / \rho_{\rm shell}$) is almost independent of the
   Lorentz factor leveling off to a value $\sim 10$ when $\Gamma_{\rm
     shell} \simgt 5$ (Fig.\,\ref{fig:variations}). As the ram
   pressure exerted by the ambient medium is proportional to $\rho
   \Gamma^2$, the pressure increase in the shocked regions rises with
   the Lorentz factor faster than the density yielding a strong
   heating of the shell's edge.  Since the ratio of the magnetic field
   and the rest mass density remains constant on both sides of the
   contact discontinuity, the magnetization parameter $\alpha_{\rm
     B,L}$ is reduced in the shocked ambient medium by up to two
   orders of magnitude for high Lorentz factors ($\Gamma_{\rm shell}
   \simgt 100$; see panels (a) and (c) in
   Fig.\,\ref{fig:variations}). This decrease of $\alpha_{\rm B,L}$ is
   important in those cases where the shells take a long time to
   collide, since the reverse shock (originating from the interaction
   with the ambient medium) can traverse a substantial part of the
   faster shell before collision.

  While shell heating also occurs for non-magnetized shells, the
  pre-collision evolution of magnetized shells is further
  characterized by an increase of the magnetization parameter
  $\sigma_{\rm L}$ in the shocked part of the shell (see
  Sec.\,\ref{shells_medium}), because $\sigma_{\rm L}$ is proportional
  to the rest-mass density when the field is oriented perpendicular to
  the direction of motion. On the other hand, the ratio of magnetic
  energy density and thermal energy density $\alpha_{\rm B,L}$
  decreases as the jump in the thermal pressure in the shocked region
  is larger than the jump in density, and thus also larger than the
  jump in the magnetic field. Consequently, at the onset of the
  collision phase the magnetic field of the faster shell is
  \emph{larger} but dynamically \emph{less} important than the shell's
  initial field.  The properties of the shocked shell (magnetization
  parameters, density contrast with respect to the shell, pressure
  jump, etc.) depend only weakly on its Lorentz factor when the latter
  becomes large. This fact is of special relevance in the case of GRB
  afterglows, which may arise from the evolution of a single shell
  moving initially with a Lorentz factor $\Gamma \sim 100$. The
  properties of the shocked shell are then largely determined by the
  magnetization parameter $\sigma$ and its density contrast with
  respect to the ambient medium.

\subsubsection{Collision phase}

  Figs.\,\ref{fig:mhdmodels} and \ref{fig:mhdmodels1} (middle column)
  demonstrate that the ratio of the magnetic field energy density
  (thin solid line) and the pressure (thick dashed line) has decreased
  at the collision phase relative to its initial value (first column),
  consistent with what is seen in Fig.\,\ref{fig:idealint} (the thin
  dotted line, denoting $\ab$, decreases in shocked regions). The
  spatial distribution of the thermal pressure in the region limited
  by the internal shocks (created during the collision) is very
  similar for all models. However, the shock structure of the slower
  shell resulting from its interaction with the ambient medium is
  characterized by a larger jump in the thermal pressure jump in
  models with larger initial magnetic fields.

  For the conditions considered in our models, which correspond to the
  typical values expected in blazars, maximum efficiencies of
  conversion of kinetic into thermal energy of $\sim 2\%$ are
  found. Much smaller values are computed for the conversion of
  kinetic into magnetic energy ($\sim 0.2\%$). According to our
  analytic model (see Appendix\,A), the maximum efficiencies of
  conversion of kinetic into thermal energy may rise up to $\sim 10\%$
  for the typical conditions expected for internal collisions in GRBs,
  while the efficiency of conversion of kinetic into magnetic energy
  decreases to $\epsmag \simlt 8 \times 10^{-5}$.
   
\subsubsection{Post-collision phase}

  Once the internal shocks break out of the merged shell, they rapidly
  stretch it, lowering its mean density and, in general, producing
  rarefaction waves in the rear part of the merged shell (similar to
  what MAMB05 found in the pure RHD case). At source frame times later
  than $\approx 1500\,$ks corresponding to a few sound crossing times
  of the merged shells the lateral expansion of the shocked regions
  can be significant, \ie the synchrotron luminosity is overestimated,
  and the size and thermodynamic properties of the merged shells have
  to be considered with care. Indeed, overestimating the pressure and
  the density in the region bounded by internal shocks could be the
  reason for some features observed at very late times in our
  synthetic light curves and in the hardness ratio (\eg the secular
  increase of the hardness ratio in the \TG{}-models).

\subsection{Evolution of the magnetic field}
   \label{concl:mag_evolution}

\begin{figure}
     \centering
     \includegraphics[scale=0.32]{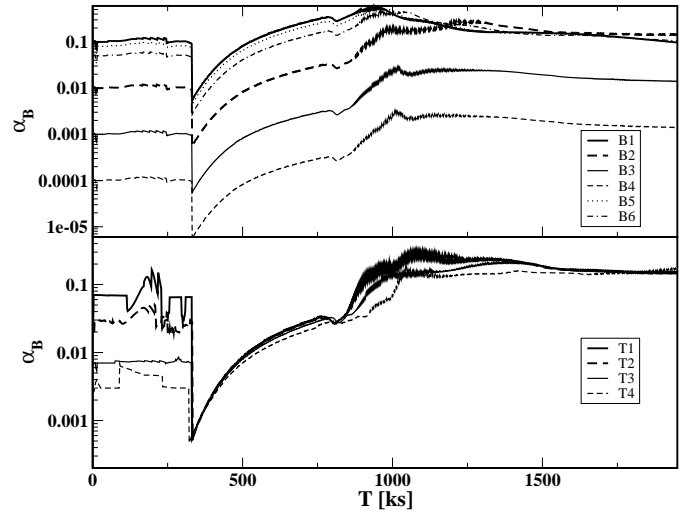}
     \caption{Ratio of the comoving magnetic energy density and the
     thermal energy density $\ab$ as a function of time in the source
     frame for the grid point with the maximum magnetic field.  The
     upper panel shows the \BG{}-models which all have the same
     initial pressure, and the lower panel displays the \TG{}-models
     which all have the same initial magnetic field. }
     \label{fig:bmax} 
   \end{figure}

   The evolution of the magnetization parameter $\ab$ at the point of
   maximum magnetic field at any given instant of time in the source
   frame shows two epochs separated by a sharp discontinuity at about
   $300\,$ks (Fig.\,\ref{fig:bmax}). The instantaneous location of the
   maximum magnetic field is indicative of the region which
   instantaneously (in the source frame) contributes most to the
   radiated energy. Hence, it is a representative point of the binary
   shell evolution.
   \begin{figure}
     \centering
     \includegraphics[scale=0.32]{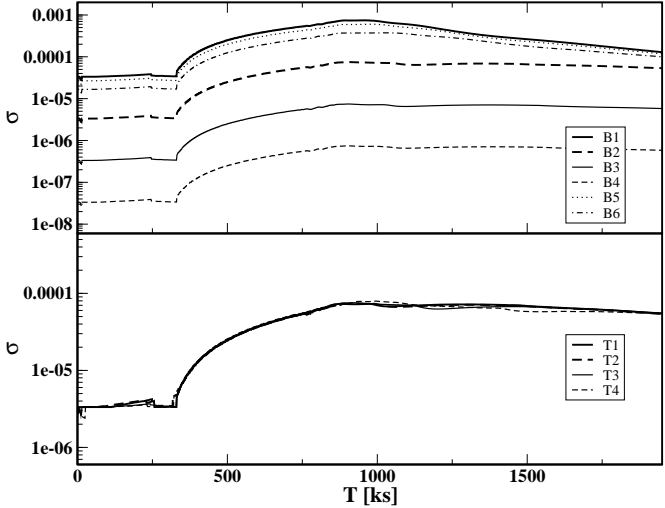}
     \caption{Same as Fig.\,\ref{fig:bmax} but showing the time
       evolution of the ratio of the comoving magnetic energy density
       and the kinetic energy density. }
     \label{fig:bmax2} 
   \end{figure}
   Left of the discontinuity the maximum magnetic field is produced by
   the reverse shock (caused by the interaction with the ambient
   medium) of the faster shell prior to the collision as it compresses
   the fluid in the shell. The part right of the discontinuity
   corresponds to a point inside the reverse internal shock formed
   during the collision phase (see Sec.\,\ref{concl:interaction}). The
   sharp discontinuity in $\ab$ occurs when the global maximum of the
   magnetic field shifts from the external shock region of the faster
   shell to the internal shock region of the merged shell.

   All curves shown in the upper panel of Fig.\,\ref{fig:bmax} are
   qualitatively very similar, especially regarding the relative
   magnitude of the sharp drop in $\ab$ at $~300\,$ks, which is almost
   identical for all the \BG{}-models. This implies that the system
   does not completely ``forget'' its initial conditions. Although
   inside the region shocked by internal shocks $\ab$ is about one
   order of magnitude smaller than the initial magnetization $\ab^0$
   in the shell, the ordering of the models according to $\ab^0$ is
   preserved, at least until $\ab$ reaches its maximum value at $t
   \sim 900\,$ks. 

   The second magnetization parameter $\sigma^0$ is {\it non
   degenerate} in the \BG{}-models (Fig.\,\ref{fig:bmax2},
   upper panel), since during the collision kinetic energy is
   dissipated, increasing the magnetic field in the shocked region,
   while the initially small thermal energy component has very little
   influence.

   A qualitatively different behavior is observed for the models of
   the \TG{}-series (Fig.\,\ref{fig:bmax}, lower panel). These have a
   different initial magnetization parameter $\ab^0$, but the
   \emph{absolute} value of the magnetic field is initially the same
   in these models. Thus, $\ab$ evolves very similarly in these models
   after the discontinuity in $\ab$ is encountered.  The evolution of
   $\sigma$ is degenerate (see lower panel of Fig.\,\ref{fig:bmax2}),
   since all models have the same initial value $\sigma^0$ and undergo
   an almost indistinguishable evolution.

   We therefore conclude that, in order to describe the essentials of
   the magnetohydrodynamic properties of binary collisions of cold
   shells, one can safely exclude the initial thermal pressure (or,
   equivalently, the internal energy) as a relevant parameter, and may
   use some canonical value instead, which is sufficiently small
   compared to the kinetic energy.

  \subsection{Radiation}
    \label{concl:radiation}
    
    The light curves of all models having the same initial magnetic
    field look similar (Fig.\,\ref{fig:LC}; right panels), which is
    consistent with the picture described in
    Sect.\,\ref{concl:mag_evolution}, as during the collision phase
    the initial value of $\sigma$ is the relevant parameter to account
    for the magnetic field evolution. For large values of $\sigma$ the
    light curves are shorter and multi-peaked, while for decreasing
    $\sigma^0$ they become single-peaked and have a longer decreasing
    tail which is progressively softer (Fig.\,\ref{fig:LC}; left
    panels). The border separating cases with a multi-peaked or
    mono-peaked light curve is given by the combination of the value
    of $\sigma$ and a critical value of $\ab^0$ of the order of
    $10^{-2}$ (see Sect.\,\ref{lc}).

    Comparing the light curves of the current work with those given in
    Fig.\,4 of MAMB05, we find that also in the magnetized case hard
    light curves are shorter in duration than soft ones, and that only
    low magnetization ($\ab^0 \leq 10^{-2}$) models have a shape
    similar to those of MAMB05, which is not surprising since in the
    limit of $\ab^0 \rightarrow 0$ the pure hydrodynamic limit must be
    recovered.  As can be seen when comparing the lower panels of
    Fig.\,\ref{fig:LC} with the lower panel of Fig.\,4 of MAMB05, the
    initial rise of the hard light curves is not smooth, but exhibits
    a small 'kink' both for non-magnetized and magnetized models. This
    similarity indicates that the assumptions made in MAMB05 are valid
    for small $\ab^0$, but need not necessarily hold for high initial
    $\ab^0$, too.

    \begin{figure}
       \centering
      \includegraphics[scale=0.3]{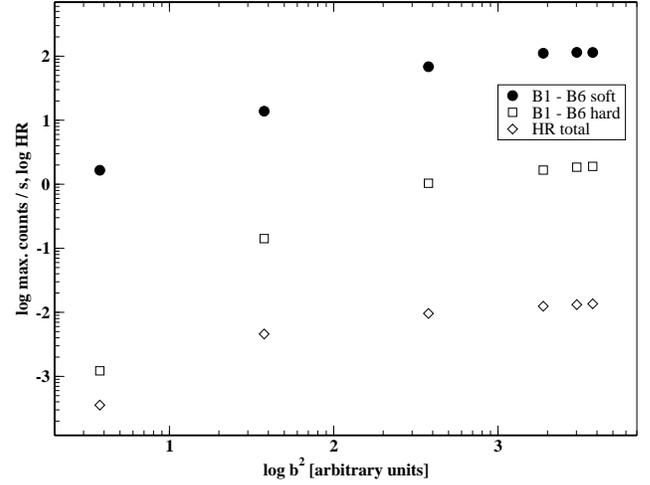}
      \caption{ Maxima of soft (filled circles) and hard (squares)
        light curves for models \BG{1} to \BG{6} as a function of the
        square of the initial comoving magnetic field. The total
        hardness ratio (HR) is defined as the ratio of the
        time-integrated number of counts in the hard and in the soft
        bands, respectively. It is displayed with the diamond symbols.}
      \label{fig:radB} 
    \end{figure}

\begin{figure}
      \centering
      \includegraphics[scale=0.32]{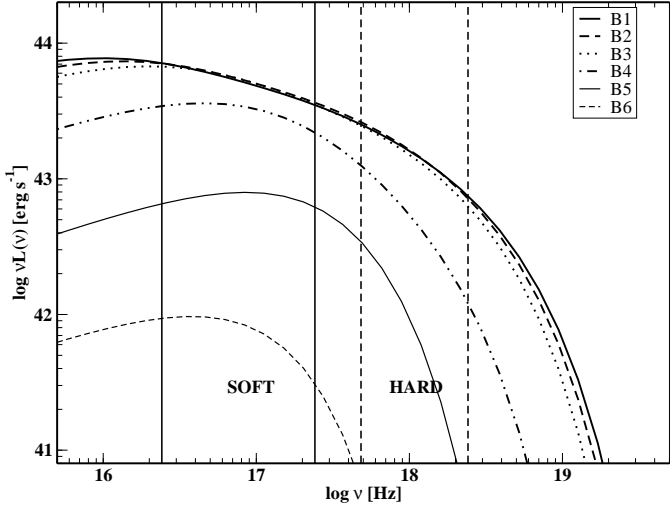}
      \caption{Spectra at the soft light curve maximum for models of
        the \BG{}-series. In order to provide an absolute
          luminosity scale we assume a shell radius
          $R_\mathrm{s}=10^{17}\,$cm. The edges of the soft and hard
        observation bands are indicated by two pairs of vertical
        lines. }
      \label{fig:specB} 
    \end{figure}

    The maxima of the soft and of the hard light curves, as well as
    the total hardness ratio (ratio of the time-integrated number of
    counts in the hard and in the soft bands) depend on the initial
    magnetic field for the \BG{}-models (Fig.\,\ref{fig:radB}). Our
    models show that the maxima of the light curves have a
    logarithmic-like dependence on the initial magnetic field energy
    density. In order to explain the weaker dependence of the light
    curve maxima and hardness ratio on the magnetic field for
    increasing magnetization, we approximate the typical high-energy
    synchrotron spectrum (Fig.\,\ref{fig:specB}) as a broad peak
    with a power-law decay towards higher photon energies. The
    position of the peak shifts to higher energies for larger values
    of the magnetic field.
    \footnote{For $\ab > 0.01$ the spectra of the models become more
      flattened in the soft band, and the global maximum shifts
      towards lower energies. This effect is partly due to the faster
      cooling of the particles, which radiate at lower energies as
      time progresses, and partly due to the fact that the spectra
      emitted by particle populations at later source times (having
      less energetic distributions) are still visible in the
      observational band (due to a sufficiently high magnetic field).
      This complicates the observed spectrum in the soft band.
      Nevertheless, the high energy part of the spectrum has basically
      the same shape as that of the low-$\ab$ models. \cite{GGC04}
      (2004) found that a similar mechanism (decrease of the peak
      spectral energy plus an steeper increase of the $\nu F_\nu$
      spectrum below the peak) may account for the larger hardness
      ratio of short GRBs. In our case, this behavior is triggered by
      the magnetic field while in \cite{JAMP06} (2006) the effect was
      attributed to the Lorentz factor.}.
    For models with lower magnetic fields the part of the spectrum in
    the hard observational band is the power-law tail resulting in a
    strong dependence of the light curve on the value of the magnetic
    field during the evolution.  For the high magnetic field models,
    the maximum of the spectrum partly extends into the hard band, \ie
    increasing the magnetic field does not change the observed energy
    in the hard band as dramatically as in the previous case. This
    can explain the saturation of the dependence of the
    hardness ratio on the magnetic field. 

    Here we stress once more that the light curve at any given time is
    the result of the sum of the emission from different points
    radiating at different times in the source frame. This explains
    why, \eg the spectrum of model \BG{1} (Fig.\,\ref{fig:specB}) has
    a more complicated shape than that of model \BG{6}. Nevertheless,
    our basic conclusion about the dependency on the magnetic field
    should hold in both cases.

    A similar study as for the \BG{}-models has been done for the
    light curve maxima of models \TG{1} to \TG{4} as a function of the
    initial thermal pressure. A very weak dependence of the radiation
    on the initial (small) pressure was found, which is consistent
    with the discussion of Sec.\,\ref{concl:mag_evolution} about the
    importance of the initial value of $\sigma$ and the unimportance
    of $\ab^0$ for the collision dynamics and the emitted radiation.

  \subsection{On the energy subtraction mechanism}
  \label{sec:subtraction}
   \begin{figure*}
      \centering
      \includegraphics[scale=0.3]{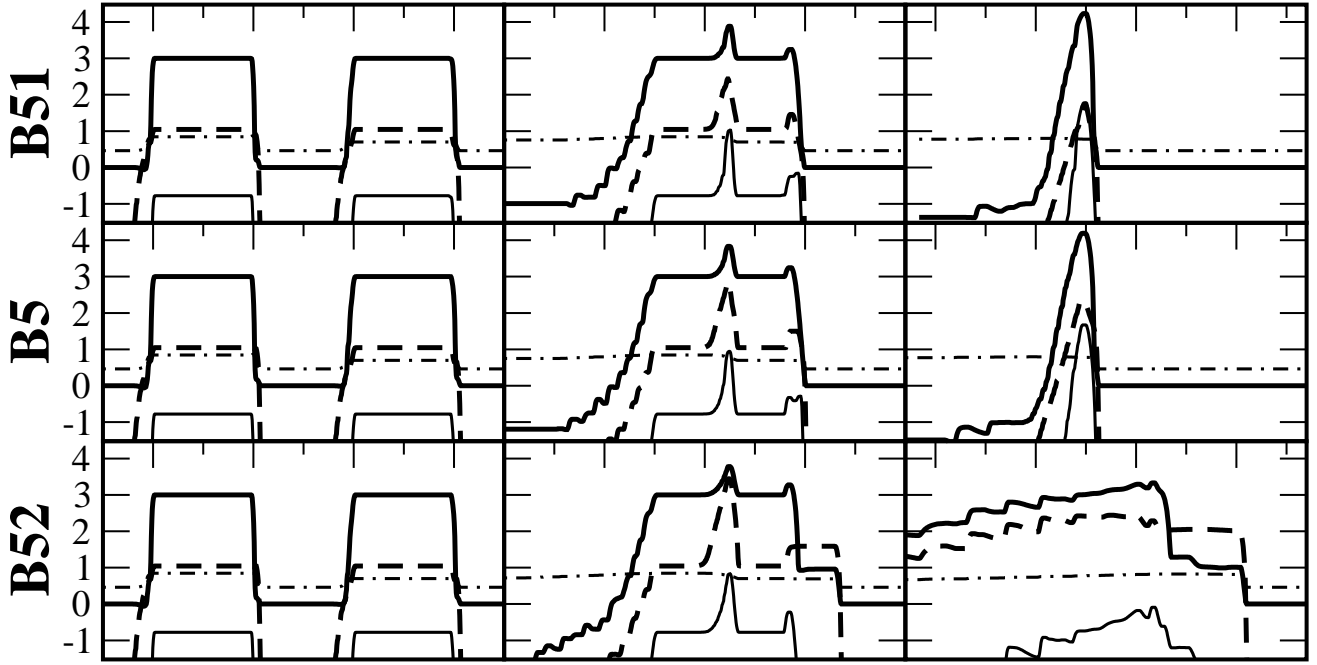}
      \caption{Same as Fig.\,\ref{fig:mhdmodels} but for model \BG{5}
        (middle column) and two additional models: model \BG{51} with
        $\ae = 0.05$ (left column), and model \BG{52} with $\ae =
        0.001$ (right column).
        }

      \label{fig:alphas} 
    \end{figure*}

    The mechanism used by us to extract energy from the thermal plasma
    and to transfer it to the non-thermal population of radiating
    electrons is based on the {\it type-E} macroscopic recipe proposed
    by MAMB04. In the type-E model one subtracts from the thermal
    plasma a fraction $\ae$ of the internal energy dissipated in
    shocks. This fraction $\ae$ is a parameter of our models which is
    poorly constrained by the current knowledge of the physics of
    particle acceleration at shock fronts, a typical value being $\ae
    \simeq 10^{-2}$.  In this subsection we critically discuss the
    impact of different choices of $\ae$ on the result. To this end we
    computed two additional models having the same model parameters as
    model \BG{5}, but assuming different values for $\ae$: In model
    \BG{51} we set $\ae = 5\times10^{-2}$, and in model \BG{52} we
    choose $\ae = 10^{-3}$ (see Fig.\,\ref{fig:alphas}).

    Regarding the physical parameters of shocked regions (\ie those
    regions affected by the energy subtraction mechanism) a comparison
    of the evolution of models \BG{5}, \BG{51} and \BG{52} provides
    the following insights:
  \begin{enumerate}

  \item Increasing the value of $\ae$ yields smaller values of the
  thermal pressure, \ie the shocked regions become cooler.
  \item The magnetic energy density grows as $\ae$ is increased in the
  region swept up by internal shocks. This growth is consistent with
  the fact that a more efficient cooling needs higher magnetic field
  strengths (because that reduces the cooling time).
  \item The size of the region containing the shocked ambient medium
  ahead of the slower shell (R-region in Fig.\,\ref{fig:flow})
  increases as $\ae$ decreases, because the amount of cooling is
  reduced (point 1 above), \ie the forward shock is not as severely
  weakened as in models with larger $\ae$.  Indeed for $\ae > 0.01$
  the forward shock is almost suppressed, and thus almost no R-region
  forms. The increase in size of the R-region leads to a decrease of
  the density, and consequently of the magnetic field energy density
  in that region (Fig.\,\ref{fig:alphas} panels in middle row). Note
  that this trend is opposite to the increase of the magnetic field
  energy density observed in the internal shock region.
  \item The state reached by models with different $\ae$ at late
  epochs is quantitatively very different (Fig.\,\ref{fig:alphas}
  panels in bottom row). The energy subtracted from the merged shell
  (which is swept up completely by shocks) is higher for increasing
  values of $\ae$. Hence, the merged shells of models \BG{5} and
  \BG{51} are much cooler (about one order of magnitude) than that of
  model \BG{52}, \ie the most evident effect of the shell cooling is
  the much smaller expansion of the merged shell at late times.
  \item The larger the value of $\ae$, the larger is the magnetization
  parameter $\ab$. In the collision phase, $500$\,ks after the start
  of simulation (Fig.\,\ref{fig:alphas} panels in middle row), the
  value of $\ab$ at the point having the maximum density (in the
  internal shock region) is more than one order of magnitude higher
  for model \BG{51} ($\ab = 3.4 \times 10^{-3}$) than for model
  \BG{52} ($\ab = 1.5 \times 10^{-4}$). The increase of $\ab$ results
  from the combined effect of the decrease of the thermal pressure
  (point 1 above) and the increase of the magnetic field strength
  (point 2 above).  This result is consistent with the physical
  expectation that larger magnetic fields (which, in our case, also
  imply larger values of $\ab$) will cool down the shocked regions
  faster, \ie for the same elapsed time, more energy will be radiated
  away for larger magnetic fields than for smaller ones.

  \item The value of $\sigma$ at the point of maximum density is
  rather insensitive to the choice of $\ae$ (for models \BG{51} and
  \BG{52} the value of $\sigma$ at $500$\,ks is $2.7 \times 10^{-6}$
  and $2.2 \times 10^{-6}$, respectively). This strengthens our
  argument that $\sigma$ is the appropriate parameter to trace the
  RMHD evolution of colliding shells, as the ordering of the models
  according to $\sigma$ remains unchanged even if one varies the value
  of $\ae$.

  \end{enumerate}

\subsection{Comparison with blazar properties}
    \label{concl:blazars}

    \subsubsection{Emission properties}
    For models \BG{1} - \BG{6} the radiative luminosity of a single
    flare lies in the range $10^{42} - 10^{44}$ erg s$^{-1}$ assuming
    a shell radius
      \footnote{Since our numerical study is one-dimensional, the
                luminosity of the models scales as $\pi
                R_\mathrm{s}^2$, where $R_\mathrm{s}$ is the shell
                radius.}
      of $10^{17}\,$cm (Fig.\,\ref{fig:specB}).  This luminosity
      roughly agrees with the lower limit observed for the moderately
      luminous blazars whose synchrotron peak lies in the X-ray band
      (see \eg \cite{FO98} 1998). As the absolute value of the
      luminosity depends non-linearly on the parameter $\ae$ (see
      \S~\ref{sec:subtraction}), the efficiency of the conversion of
      kinetic energy into radiation is somewhat unconstrained. For a
      homogeneous cylindrical shell with a rest mass density $\rho$,
      Lorentz factor $\Gamma$, and velocity $v$ the kinetic luminosity
      can be defined as
      \[ L_{\rm kin} := \pi R_\mathrm{s}^2 (\tilde{h}\Gamma -1)\rho 
                        \Gamma v\,. 
      \]
      For our initial models the kinetic luminosity is $\simeq 2\times
      10^{47}$ erg s$^{-1}$, implying a conversion efficiency of the
      order of $0.01\% - 0.1\%$ for $\ae=10^{-2}$. We emphasize that
      although both the kinetic and the radiative luminosities are
      unconstrained by our one-dimensional simulations up to a factor
      $\pi R_\mathrm{s}^2$, their ratio and hence the efficiency of
      conversion of kinetic into radiative luminosities is independent
      of this factor.

      Our models are set up to have a uniform magnetization $\sigma^0$
      both in the shells and in the ambient medium. Consequently, once
      the density of the ambient medium is fixed, the magnetic field
      strength in the shocked region\footnote{The absolute scale of
        $B$ depends on $\rho_{\rm ext}$ (Sec.\,\ref{initial}).  Since
        the value of the magnetic field strength in the shocked region
        ($B_{\rm shock}$) may be computed either analytically or
        numerically, we can tune $\rho_{\rm ext}$ such that $B_{\rm
          shock}\simeq 0.1 - 1$ G.}  roughly corresponds to the values
      inferred from observations of blazars, \ie $\simeq 0.01 - 1$ G
      (\eg \cite{GH98} 1998). Increasing the ratio of shell to ambient
      medium density $\chi_\rho$, will increase the kinetic luminosity
      of the shells proportionally.  Assuming a uniform initial
      magnetization, an increase of $\chi_\rho$ will, however, drive
      the value of the magnetic field in the shocked regions above the
      aforementioned values. Thus, in order to set up future
      simulations that mimic more powerful blazars with realistic
      magnetic field strengths, we will have to consider denser shells
      whose magnetization parameter $\sigma^0$ is smaller than that of
      the ambient fluid.

      The number density of radiating non-thermal electrons is $\simgt
      10^4 - 10^5\,$cm$^{-3}$, which is one or two orders of magnitude
      higher than the usually inferred values $\simgt 10^3 -
      10^4\,$cm$^{-3}$ (e.g. \cite{GH98} 1998). However, these values
      are obtained under the assumption that the emitting region is a
      sphere, whereas in our case the emitting region is a cylinder
      whose height $L\sim 10^{14}\,$cm is much smaller than its radius
      $R_\mathrm{s}\sim 10^{17}\,$cm. Hence, for the same number of
      emitting electrons, the number density of radiating particles in
      this smaller volume is expected to be larger.

    \subsubsection{Light curves}
    Following \cite{BR05}\,(2005), we calculate the cross-correlation
    function (CCF) of the light curves of models \BG{1} to \BG{6}. If
    the observed counts are binned into $N$ equidistant time intervals
    $\Delta t$, the CCF is defined as follows,
% CCF
    \begin{equation}
      CCF(k\Delta t) := \dsfrac{\sum_i (x_{\rm soft}(i\Delta t) - \bar{x}_{\rm soft})
                (x_{\rm hard}((i+k)\Delta t) - \bar{x}_{\rm hard})}{n(k\Delta t)
               \sqrt{\sigma_{\rm soft}^2\sigma_{\rm hard}^2}}\, ,
    \label{CCF}
    \end{equation}
% end of CCF 
    where $k=0, 1, ..., (N-1)$, and $\bar{x}_{\rm soft}$, and
    $\bar{x}_{\rm hard}$ are the number of counts averaged over the
    N-intervals for photons detected in the soft and in the hard band,
    respectively.  $\sigma_{\rm soft}$ and $\sigma_{\rm hard}$ are the
    standard deviations of the samples with respect to the
    corresponding average number of counts in the soft and hard bands,
    respectively. $n(k)$ is the number of pairs $(i, i+k)$ where
    $x_{\rm soft}(i\Delta t)$ and $x_{\rm hard}((i+k)\Delta t)$ are
    both nonzero for a given $k$.  The time interval $k
    \Delta t$ is called the time lag. Significant correlation at \eg
    negative lags implies that soft band variations are observed later
    than hard band variations.

    \begin{figure}
      \centering
      \includegraphics[scale=0.32]{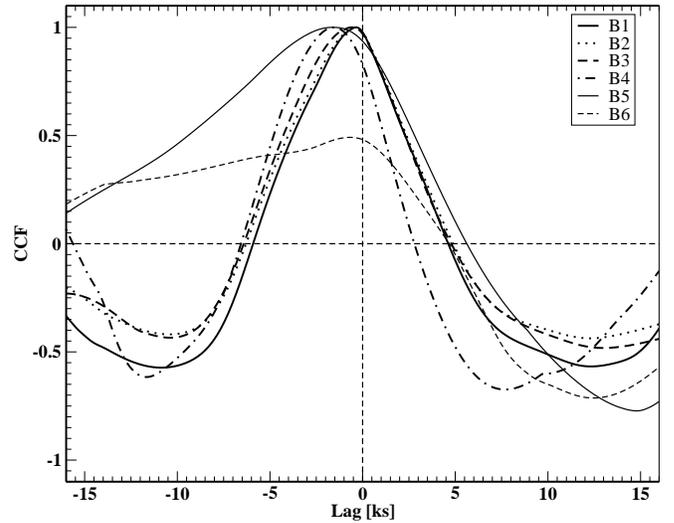}
      \caption{ Normalized cross-correlation function (see
        Eq.\,\ref{CCF}) for models \BG{1} to \BG{6}. Bins of $16.6\,$s
        were used when computing the cross-correlation functions.  }
      \label{fig:CCF} 
    \end{figure}

    Contrary to the observations shown by \cite{BR05} (2005; Fig.\,3),
    the maxima of the CCFs are not located at zero time lag, but at
    negative time lags (Fig.\,\ref{fig:CCF}), \ie our simulations only
    predict the existence of {\it soft lags}. However, the overall
    shapes of the CCF curves are well covered by our models, including
    both symmetric and asymmetric ones.

    For small initial magnetic fields, the CCFs become more
    asymmetric. \cite{R04}\,(2004) pointed out that the asymmetry can
    be reproduced by modeling light curves as initially rising
    linearly, and subsequently decreasing with different time scales
    at different frequencies. This behavior can be understood from our
    models as well. For lower magnetic fields the light curve is
    visible for a shorter time in the hard band than in the soft band
    (Fig.\,\ref{fig:specB}), thus giving rise to different time
    scales. We point out that these time scales are not only related
    to the cooling of the relativistic electrons, but to the radial
    expansion of the emission region as well, whereby the magnetic
    field inside this region is decreased. The fact that we only see
    soft lags is probably a consequence of the particular acceleration
    mechanism model we have chosen (fixed lower and upper limits of
    the non-thermal electron energy distribution; type-E model of
    MAMB04).  The influence of the acceleration mechanism will be the
    subject of future work.

\subsection{Summary}
   
    A detailed study of the interaction of magnetized shells in
    relativistic outflows has been performed. An idealized interaction
    of a shell with an ambient medium was studied using an exact
    Riemann solver (\cite{RO05} 2005), while the full interaction of
    two colliding shells was simulated in one spatial dimension using
    a RMHD code (\cite{LA05} 2005) coupled to the synchrotron emission
    scheme of MAMB04. The models are parametrized to address blazar
    jets under the hypothesis that blazar light curves result from
    internal shocks within relativistic jets. However, some results
    can be extrapolated to the dynamics of internal shocks in GRBs,
    when taking into account that in the latter case the Lorentz
    factors are one order of magnitude or more larger than in blazar
    jets, and that therefore much smaller relative velocities are
    encountered between the colliding shells.

\begin{acknowledgements}
  The authors are grateful to the anonymous referee for her/his
  constructive comments on this work. Furthermore, the authors thank
  W.\,Brinkmann for helpful discussions and J.M. Mart\'{\i} for his
  careful reading and valuable comments. All computations were
  performed on the IBM-Regatta system of the Rechenzentrum Garching of
  the Max-Planck-Society. PM is currently staying at the University of
  Valencia with a European Union Marie Curie Incoming International
  Fellowship (MEIF-CT-2005-009395). He also acknowledges support from
  the Max-Planck-Institut f\"ur Astrophysik and the
  Max-Planck-Institut f\"ur extraterrestrische Physik. MAA is a
  Ram\'on y Cajal Fellow of the Spanish Ministry of Education and
  Science, and also acknowledges partial support by the Spanish
  Ministerio de Ciencia y Tecnolog\'{\i}a (AYA2004-0067-C03-C01).
\end{acknowledgements}

\begin{appendix}

\section{Analytic modeling of energy conversion efficiencies}
\label{app:eff}

\begin{figure}
  \centering
  \includegraphics[scale=0.5]{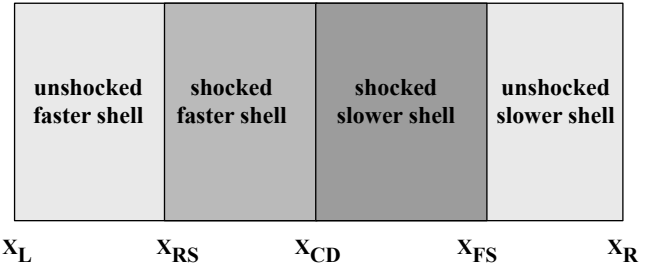}
  \caption{A schematic representation of the idealized interaction of
  two colliding shells consisting of four distinct regions. The shells
  are assumed to move towards the right. }
  \label{fig:app_int} 
\end{figure}

In this section we idealize the interaction of two colliding shells by
neglecting the pre-collision phase, and assuming that the dynamics of
the collision can be approximated by solving a Riemann problem at the
interface of the two shells. Fig.\,\ref{fig:app_int} shows an example
of the flow structure in this case. From left to right we identify the
following four distinct regions
\footnote{The fluid in the faster and slower shocked shell regions has
the same Lorentz factor as the contact discontinuity, \ie
$\Gamma_\mathrm{CD}$.},
each with its rest-mass density $\rho$, thermal pressure $p$, magnetic
field $B$ (assumed perpendicular to the direction of propagation), and
the fluid Lorentz factor $\Gamma$:
\begin{enumerate}
  \item unshocked faster shell 
   (between $x_\mathrm{L}(T)$ and $x_\mathrm{RS}(T)$): $\rho_\mathrm{L}$, $p_\mathrm{L}$, $B_\mathrm{L}$, $\Gamma_\mathrm{L}$
  \item shocked faster shell 
   (between $x_\mathrm{L}(RS)$ and $x_\mathrm{CD}(T)$): $\rho_\mathrm{LS}$, $p_\mathrm{LS}$, $B_\mathrm{LS}$, $\Gamma_\mathrm{CD}$
  \item shocked slower shell 
   (between $x_\mathrm{CD}(T)$ and $x_\mathrm{FS}(T)$): $\rho_\mathrm{RS}$, $p_\mathrm{RS}$, $B_\mathrm{RS}$, $\Gamma_\mathrm{CD}$
  \item unshocked slower shell 
    (between $x_\mathrm{FS}(T)$ and $x_\mathrm{R}(T)$): $\rho_\mathrm{R}$, $p_\mathrm{R}$, $B_\mathrm{R}$, $\Gamma_\mathrm{R}$
\end{enumerate}
The five (time-dependent) boundaries of the four regions are from left
to right:
\begin{enumerate}
  \item left edge of the faster shell at 
    $x_\mathrm{L}(T)  = T\sqrt{1-\Gamma_\mathrm{L}^{-2}}$
  \item reverse shock at
    $x_\mathrm{RS}(T) = \Delta x_\mathrm{L} + T\sqrt{1-\Gamma_\mathrm{RS}^{-2}}$
  \item contact discontinuity at 
    $x_\mathrm{CD}(T) = \Delta x_\mathrm{L} + T\sqrt{1-\Gamma_\mathrm{CD}^{-2}}$
  \item forward shock at
    $x_\mathrm{FS}(T) = \Delta x_\mathrm{L} + T\sqrt{1-\Gamma_\mathrm{FS}^{-2}}$
  \item right edge of the slower shell at 
    $x_\mathrm{R}(T)  = \Delta x_\mathrm{L} + \Delta x_\mathrm{R} + T\sqrt{1-\Gamma_{R}^{-2}}$,
\end{enumerate}
where $\Delta x_\mathrm{L}$ and $\Delta x_\mathrm{R}$ are the initial
thickness of the faster and of the slower shell, respectively. The
reverse shock, the contact discontinuity and the forward shock
originate at the interface between the two shells. The origin of the
coordinate system coincides with the left edge of the faster shell at
$T=0$.

Knowing the initial values of $\rho$, $p$, $B$, and $\Gamma$ in the
shells, we use the exact Riemann solver (\cite{RO05} 2005) to compute
values in the intermediate states. We then define the total kinetic,
thermal and magnetic energy at a time $T$ as follows:
\begin{equation}\label{eq:app_ekin}
  \begin{array}{rcl}
    E_\mathrm{K}(T) & :=& \Gamma_\mathrm{L}(\Gamma_\mathrm{L}-1)\ \rho_\mathrm{L}\ (x_\mathrm{RS}(T) - x_\mathrm{L}(T)) \\
    & & + \Gamma_\mathrm{CD}(\Gamma_\mathrm{CD}-1)\ \rho_\mathrm{LS}\ (x_\mathrm{CD}(T) - x_\mathrm{RS}(T)) \\
    & & + \Gamma_\mathrm{CD}(\Gamma_\mathrm{CD}-1)\ \rho_\mathrm{RS}\ (x_\mathrm{FS}(T) - x_\mathrm{CD}(T)) \\
    & & + \Gamma_{R}(\Gamma_{R}-1)\ \rho_\mathrm{R}\ (x_\mathrm{R}(T) - x_\mathrm{FS}(T))\, 
  \end{array}\, ,
\end{equation}
\begin{equation}\label{eq:app_eth}
  \begin{array}{rcl}
    E_\mathrm{th}(T) & :=& \left(\dsfrac{\gad}{\gad-1}\Gamma_\mathrm{L}^2 -1 \right)\ p_\mathrm{L}\ (x_\mathrm{RS}(T) - x_\mathrm{L}(T)) \\
    & & + \left(\dsfrac{\gad}{\gad-1}\Gamma_\mathrm{CD}^2 -1 \right) p_\mathrm{LS}\ (x_\mathrm{CD}(T) - x_\mathrm{RS}(T)) \\
    & & + \left(\dsfrac{\gad}{\gad-1}\Gamma_\mathrm{CD}^2 -1 \right) p_\mathrm{RS}\ (x_\mathrm{FS}(T) - x_\mathrm{CD}(T)) \\
    & & + \left(\dsfrac{\gad}{\gad-1}\Gamma_\mathrm{R}^2 -1 \right)    p_\mathrm{R}\ (x_\mathrm{R}(T) - x_\mathrm{FS}(T))
  \end{array}\, ,
\end{equation}
and
\begin{equation}\label{eq:app_emag}
  \begin{array}{rcl}
    E_\mathrm{mag}(T) & :=& \left(\Gamma_\mathrm{L}^2 -\dsfrac{1}{2} \right)\ \dsfrac{B_\mathrm{L}^2}{4\pi}\ (x_\mathrm{RS}(T) - x_\mathrm{L}(T)) \\
    & & + \left(\Gamma_\mathrm{CD}^2 -\dsfrac{1}{2} \right)\ \dsfrac{B_\mathrm{LS}^2}{4\pi}\ (x_\mathrm{CD}(T) - x_\mathrm{RS}(T)) \\
    & & + \left(\Gamma_\mathrm{CD}^2 -\dsfrac{1}{2} \right)\ \dsfrac{B_\mathrm{RS}^2}{4\pi}\ (x_\mathrm{FS}(T) - x_\mathrm{CD}(T)) \\
    & & + \left(\Gamma_\mathrm{R}^2 -\dsfrac{1}{2} \right)\ \dsfrac{B_\mathrm{R}^2}{4\pi}\ (x_\mathrm{R}(T) - x_\mathrm{FS}(T)) \\
  \end{array}\, .
\end{equation}

As in Eq.\,\ref{eq:eps}, we define the total efficiency for
conversion of kinetic into thermal and magnetic energy
\begin{equation}\label{eq:app_eps}
  \epsilon^\mathrm{an}(T) := 1 - \dsfrac{E_\mathrm{K}(T)}{E_\mathrm{K}(0)}\, ,
\end{equation}
the efficiency of conversion of kinetic into thermal energy
\begin{equation}\label{eq:app_eps_th}
  \epsth^\mathrm{an}(T) := \dsfrac{E_\mathrm{th}(T) - E_\mathrm{th}(0)}{E_\mathrm{K}(0)}\, ,
\end{equation}
and the efficiency of conversion of kinetic into magnetic energy
\begin{equation}\label{eq:app_eps_mag}
  \epsmag^\mathrm{an}(T) := \dsfrac{E_\mathrm{mag}(T) - E_\mathrm{mag}(0)}{E_\mathrm{K}(0)}\, .
\end{equation}

The gray thick line in Fig.\,\ref{fig:eff1} shows $\epsth^\mathrm{an}$
(upper panel) and $\epsmag^\mathrm{an}$ (lower panel) for the initial
shell set-up of model \BG{1}. We assume that the shells collide after
a time of $335\,$ks, determined by their initial separation
($10^{14}\,$cm) and Lorentz factors ($7$ and $5$, respectively). We
compute the efficiencies until the forward shock reaches the edge of
the slower shell, as the analytic approach is no longer valid later
on.  There is a good agreement between the final value of
$\epsth^\mathrm{an}$ ($0.0191$) and the maximum value of $\epsth$ for
model \BG{1} ($0.0195$). The slope of the initial efficiency increase
is rather well captured, too. The instantaneous discrepancies between
the numeric and analytic estimates mostly arise from the pre-collision
evolution of the shells.

\begin{figure}
  \centering
  \includegraphics[scale=0.32]{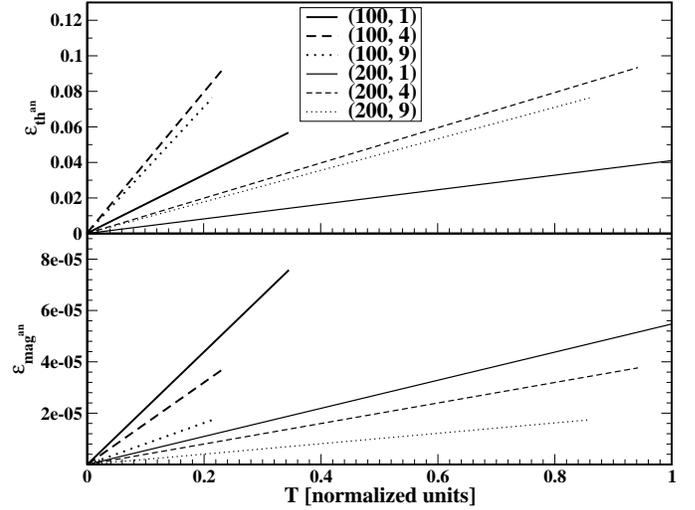}
  \caption{Efficiency of converting bulk kinetic energy into thermal
    (upper panel) and magnetic field energy (lower panel) as a
    function of time. Shown are six models with
    different values of $(\Gamma_\mathrm{R}, \Delta g)$ and with
    $\sigma=3\times 10^{-5}$, where $\Delta g \equiv
    (\Gamma_\mathrm{L} - \Gamma_\mathrm{R}) / \Gamma_\mathrm{R}$ is
    the relative Lorentz factor, and where $\Gamma_\mathrm{L}$ and
    $\Gamma_\mathrm{R}$ are the Lorentz factors of the faster and the
    slower shell, respectively.}
  \label{fig:app_grb}
\end{figure}

We have also applied the above method to the conditions expected to be
found in GRBs. Defining the relative Lorentz factor of the shells as
$\Delta g \equiv (\Gamma_\mathrm{L} - \Gamma_\mathrm{R}) /
\Gamma_\mathrm{R}$, we show $\epsth^\mathrm{an}$ and
$\epsmag^\mathrm{an}$ in Fig.\,\ref{fig:app_grb} assuming a low
magnetization parameter $\sigma=3\times 10^{-5}$ and initially cold
shells with $p/\rho = 10^{-2}$. Different curves correspond to
different values of $(\Gamma_\mathrm{R}, \Delta g)$. Our results show
that, for the typical conditions considered for internal shocks in
GRBs (\eg \cite{DM98} 1998), maximum efficiencies between $6\%$ and
$10\%$ can be obtained.

\begin{figure}
  \centering
  \includegraphics[scale=0.32]{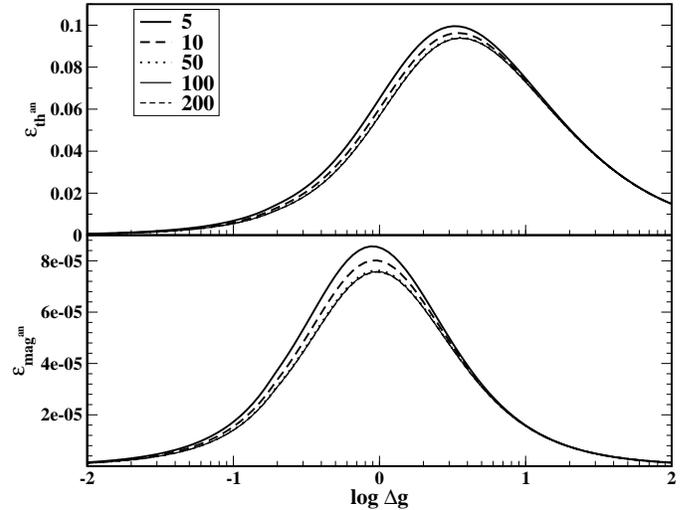}
  \caption{Efficiency of converting bulk kinetic energy into thermal
    (upper panel) and magnetic field energy (lower panel) as a
    function of the relative Lorentz factor $\Delta g$, where $\Delta
    g \equiv (\Gamma_\mathrm{L} - \Gamma_\mathrm{R}) /
    \Gamma_\mathrm{R}$, and where $\Gamma_\mathrm{L}$ and
    $\Gamma_\mathrm{R}$ are the Lorentz factors of the faster and the
    slower shell, respectively . Shown are five models with different
    $\Gamma_\mathrm{R}$ ($5$, $10$, $50$, $100$, $200$) and with
    $\sigma=3\times 10^{-5}$.}
  \label{fig:app_dg}
\end{figure}

Figure\,\ref{fig:app_dg} shows the maximum efficiencies (corresponding
to the final states in Fig.\,\ref{fig:app_grb}) for a fixed Lorentz
factor $\Gamma_\mathrm{R}$ as a function of the relative Lorentz
factor $\Delta g$ keeping all other parameters the same as in
Fig.\,\ref{fig:app_grb}. The efficiency depends strongly on $\Delta
g$, and is independent of $\Gamma_\mathrm{R}$ for large values of
$\Delta g$.
%%%
% Please check the above statement!  >>> OK
%%%

% 

We have also investigated the dependency of the efficiency of
conversion of kinetic energy into thermal and into magnetic energy on
the magnetization parameter $\sigma$ for two selected sets of the
parameters $(\Gamma_\mathrm{R}, \Delta g)$ which are representative
for blazars $(\Gamma_\mathrm{R}, \Delta g)=(5,2)$ and GRBs
$(\Gamma_\mathrm{R}, \Delta g)=(100,1)$, respectively. As we see in
Fig.\,\ref{fig:app_dg} increasing the magnetization tends to increase
$\epsilon_\mathrm{mag}^\mathrm{an}$ at all times independently of the
value of the shells' Lorentz factor (\ie both for blazars and for
GRBs). Indeed, $\epsilon_\mathrm{mag}^\mathrm{an}$ reaches values of
$0.2-0.3$ for magnetically dominated shells ($\sigma = 10$). More
moderate magnetizations ($\sigma=1$) but still much larger than the
ones we have considered for the numerical models in this work yield
$\epsilon_\mathrm{mag}^\mathrm{an} \sim 0.1$. The trends for
$\epsilon_\mathrm{th}^\mathrm{an}$ are not monotonic, and depend on
the chosen scenario. For blazar conditions,
$\epsilon_\mathrm{th}^\mathrm{an}$ decreases with increasing values of
$\sigma$. For the conditions met in GRB flows, there is an asymptotic
trend to increase $\epsilon_\mathrm{th}^\mathrm{an}$ for very large
values of the magnetization parameter ($\sigma \simgt 1$). When
$\sigma$ is increased, the efficiency of conversion of kinetic into
magnetic energy becomes larger than the efficiency of converting
kinetic into thermal energy. We point out that values of $\sigma \simgt 1$ imply that
initially the shells may have a magnetic energy larger than the
thermal energy. When the initial internal energy is larger than the
magnetic energy, then $\epsilon_\mathrm{th}^\mathrm{an} >
\epsilon_\mathrm{mag}^\mathrm{an}$ holds , while in the opposite case
$\epsilon_\mathrm{th}^\mathrm{an} >
\epsilon_\mathrm{mag}^\mathrm{an}$.

\begin{figure}
  \centering
  \includegraphics[scale=0.32]{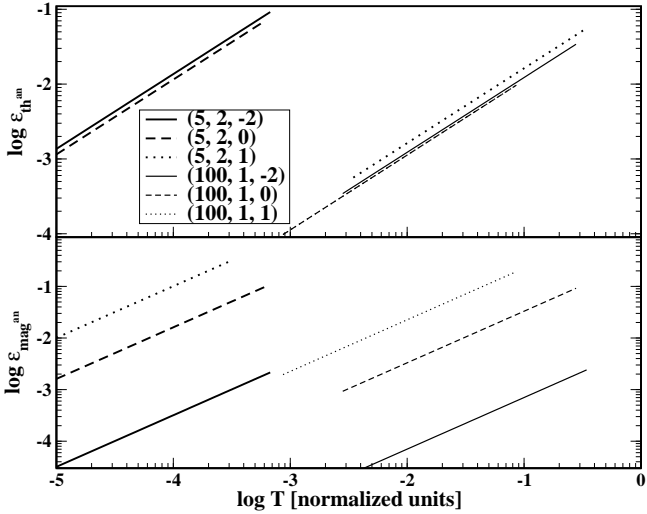}
  \caption{Same as Fig.\,\ref{fig:app_grb}, but for different values
           of $(\Gamma_\mathrm{R}, \Delta g, \log \sigma)$.}
  \label{fig:app_mag}
\end{figure}

\end{appendix}

\end{document}